\documentclass[twocolumn,aps,prb,superscriptaddress,amsmath,amssymb,superscriptaddress]{revtex4-1}
\newcommand{\pagenumbaa}{1}
\usepackage{graphicx}
\usepackage{color}
\usepackage{braket}

\usepackage{verbatim}
\usepackage{amsmath}
\usepackage{amsfonts}
\usepackage{hyperref}
\usepackage{mathtools}
\DeclarePairedDelimiter\abs{\lvert}{\rvert}
\usepackage{float}
\usepackage{lipsum}
\definecolor{orange}{rgb}{1,0.5,0}

\begin{document}

\title{Excitons in InGaAs Quantum Dots without Electron Wetting Layer States}

\author{Matthias C. L\"obl}
\email[]{matthias.loebl@unibas.ch}
\affiliation{Department of Physics, University of Basel, Klingelbergstrasse 82, CH-4056 Basel, Switzerland}
\author{Sven Scholz}
\affiliation{Lehrstuhl f\"ur Angewandte Festk\"orperphysik, Ruhr-Universit\"at Bochum, DE-44780 Bochum, Germany}
\author{Immo S\"ollner}
\affiliation{Department of Physics, University of Basel, Klingelbergstrasse 82, CH-4056 Basel, Switzerland}
\author{Julian Ritzmann}
\affiliation{Lehrstuhl f\"ur Angewandte Festk\"orperphysik, Ruhr-Universit\"at Bochum, DE-44780 Bochum, Germany}
\author{Thibaud Denneulin}
\affiliation{Peter Gr\"unberg Institute, Forschungszentrum J\"ulich, DE-52425 J\"ulich, Germany}
\author{Andras Kovacs}
\affiliation{Peter Gr\"unberg Institute, Forschungszentrum J\"ulich, DE-52425 J\"ulich, Germany}
\author{Beata E. Kardyna\l}
\affiliation{Peter Gr\"unberg Institute, Forschungszentrum J\"ulich, DE-52425 J\"ulich, Germany}
\author{Andreas D. Wieck}
\affiliation{Lehrstuhl f\"ur Angewandte Festk\"orperphysik, Ruhr-Universit\"at Bochum, DE-44780 Bochum, Germany}
\author{Arne Ludwig}
\affiliation{Lehrstuhl f\"ur Angewandte Festk\"orperphysik, Ruhr-Universit\"at Bochum, DE-44780 Bochum, Germany}
\author{Richard J. Warburton}
\affiliation{Department of Physics, University of Basel, Klingelbergstrasse 82, CH-4056 Basel, Switzerland}

\begin{abstract}
The Stranski-Krastanov (SK) growth-mode facilitates the self-assembly of quantum dots (QDs) using lattice-mismatched semiconductors, for instance InAs and GaAs. SK QDs are defect-free and can be embedded in heterostructures and nano-engineered devices. InAs QDs are excellent photon emitters: QD-excitons, electron-hole bound pairs, are exploited as emitters of high quality single photons for quantum communication. One significant drawback of the SK-mode is the wetting layer (WL). The WL results in a continuum rather close in energy to the QD-confined-states. The WL-states lead to unwanted scattering and dephasing processes of QD-excitons. Here, we report that a slight modification to the SK-growth-protocol of InAs on GaAs -- we add a monolayer of AlAs following InAs QD formation -- results in a radical change to the QD-excitons. Extensive characterisation demonstrates that this additional layer eliminates the WL-continuum for electrons enabling the creation of highly charged excitons where up to six electrons occupy the same QD. Single QDs grown with this protocol exhibit optical linewidths matching those of the very best SK QDs making them an attractive alternative to standard InGaAs QDs.
\end{abstract}

\maketitle

\setcounter{page}{\pagenumbaa}
\thispagestyle{plain}
InGaAs quantum dots (QDs) grown in the Stranski-Krastanov (SK) mode are excellent photon emitters. Individual QDs provide a source of highly indistinguishable single photons \cite{Somaschi2016,Ding2016,Kirsanske2017} and a platform for spin-photon and spin-spin entanglement \cite{Delteil2016,Gao2012,Stockill2017}. Their solid state nature enables the integration of QDs in on-chip nanostructures such as photonic crystal cavities or waveguides \cite{Lodahl2015}. In some respects, a QD can be considered as an artificial atom. However, this approximation is often too simplistic. Unlike a real atom in free space, an exciton (a bound electron-hole pair) in a QD can couple to further degrees of freedom in its solid state environment, for instance phonons \cite{Ramsay2010,Wei2014,Kaldewey2017} and nuclear spins \cite{Greilich2007,Press2008b,Prechtel2016,Majcher2017,Huthmacher2018}. One problematic source of unwanted coupling is the so-called wetting layer (WL) \cite{Karrai2003,Wang2005,VanHattem2013,Rai2016}. The WL is an inherent feature of SK-growth and represents a two-dimensional layer lying between all QDs.

On account of the confinement in the growth direction, there is an energy gap between the WL-continuum and QD-electron and QD-hole states. However, this gap protects the QD-electrons and -holes from coupling to the WL only to a certain extent. The gap vanishes for a QD containing several electrons due to the on-site Coulomb repulsions; the energy gap can be bridged by carrier-carrier and carrier-phonon scattering \cite{Karrai2003,Carmele2010}. Furthermore, the gap is not complete: a low energy tail of the WL-continuum can extend to the QD-confined-states \cite{Toda1999}. The result is that the WL has negative consequences for quantum applications. Specific examples are well known. Multi-electron states of a QD hybridize with extended states of the WL \cite{Karrai2003,Schulhauser2004,VanHattem2013} severely limiting the prospects for using multi-electron states as qubits \cite{Weiss2013}. QD--WL Auger processes can lead to a parasitic coupling between a QD and an off-resonant cavity \cite{Kaniber2008,Winger2009,Settnes2013}. A broad absorption background due to WL-states \cite{ Toda1999} leads to damping of exciton Rabi oscillations \cite{Villas2005,Wang2005} and enhanced exciton-phonon scattering \cite{Urbaszek2004}.

We show here that the QD-properties can be radically altered when WL-states are absent. Electron WL-states are removed by a simple modification to the SK-growth: InGaAs QDs are overgrown with a monolayer of AlAs \cite{Arzberger1999,Tsatsulnikov2000,Lu2015,Tutu2013}. The absence of electron WL-states for AlAs-capped QDs is explained on the nano-scale: AlAs increases the bandgap of the material laterally surrounding a QD thereby eliminating bound electron WL-states. Changes regarding the QD-properties are drastic: we observe highly charged excitons with narrow optical emission where up to six electrons occupy the conduction band shells of the QD -- a novelty for QDs in the considered wavelength regime. The QD-potential is deepened and hybridization with any WL-continuum is absent. Furthermore, the QDs have close to transform limited optical linewidths at low temperature, a very sensitive probe of the material quality \cite{Kuhlmann2013}. We propose that conventional SK QDs can be profitably replaced with their no-electron WL-counterparts in all SK QD-quantum-devices.

\section{Sample Growth and Ensemble Measurements}
\label{sec:growth}

\begin{figure*}[t]
\includegraphics[width=1.8\columnwidth]{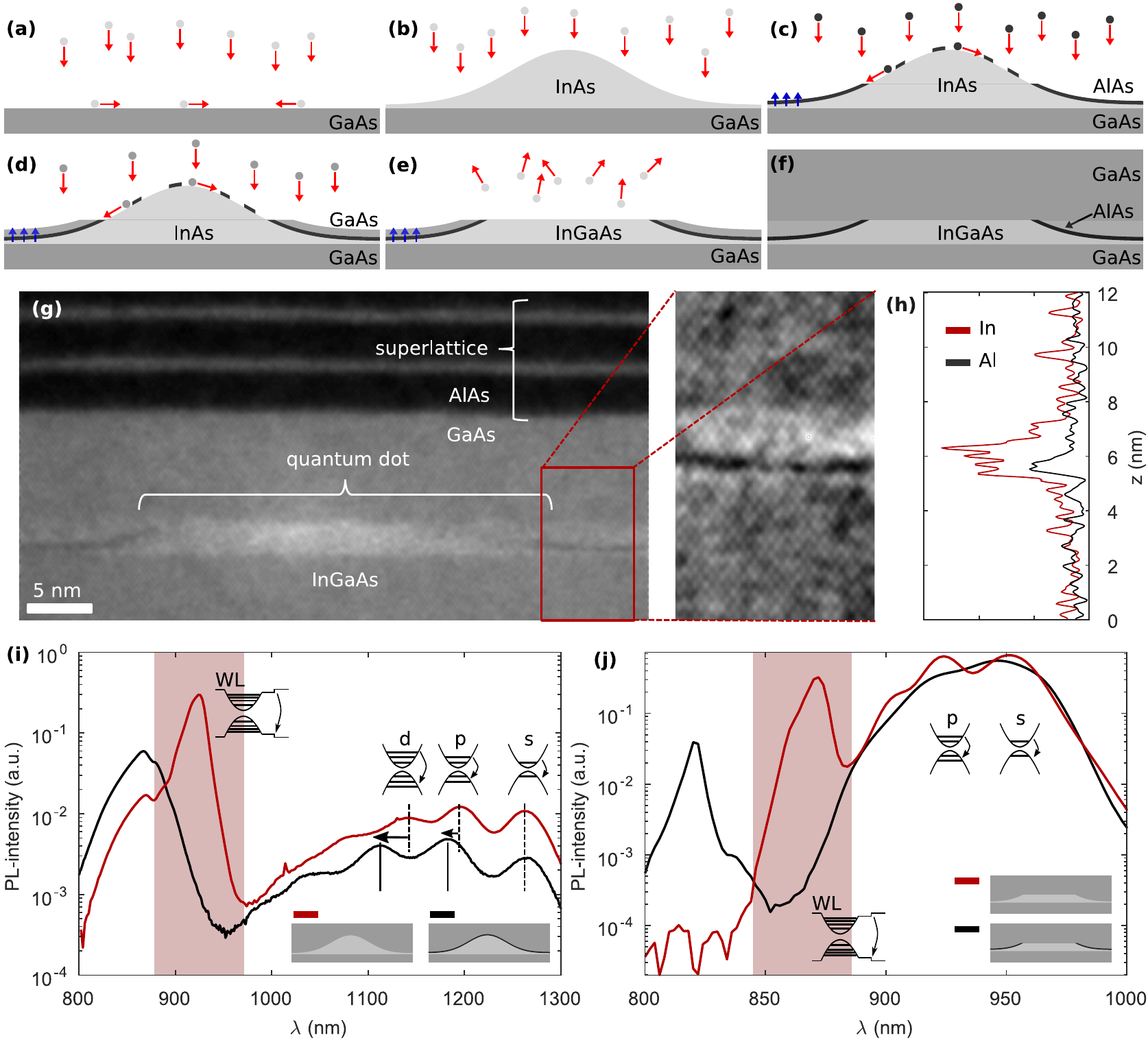}
\caption{{\bf Electron wetting layer-free quantum dots: growth and characterization.} (a-f) Schematic process of the quantum dot (QD) growth. (a) InAs is deposited on a GaAs-surface. (b) After deposition of  $\sim 1.5$ monolayers of InAs, strain-driven QD-formation takes place. The QDs are capped with a monolayer of AlAs (c) and $2\ \text{nm}$ GaAs (d). During these steps segregation of In atoms takes place (blue arrows) resulting in a wetting layer (WL) which is an alloy of GaAs, InAs, and AlAs. (e) The top part of the capped QD evaporates at 630 $^{\circ}\text{C}$ (flushing step). (f) The flushed QDs are overgrown with GaAs. (g) HAADF STEM image of a flushed QD. InAs appears bright, AlAs dark. (h) Chemical composition of the WL measured by spatially resolved energy dispersive X-ray (EDX) spectroscopy at a location without a QD (different location to (g) but nominally the same). (i) Ensemble photoluminescence (PL) at room temperature from a sample with unflushed, standard InGaAs QDs (red curve) and unflushed, AlAs-capped InGaAs QDs (black curve). The WL PL is highlighted by the red band. QD PL appears in the regime $1,000-1,300\ \text{nm}$. The QD-shells are labelled. (j) Ensemble-PL at $77\ \text{K}$ from a sample with flushed, standard InGaAs QDs (red curve) and flushed, AlAs-capped InGaAs QDs (black curve). The flushing blue-shifts the QD-ensemble to $\sim900-980\ \text{nm}$.}
\label{fig:growth}
\end{figure*}

The QDs are grown by molecular beam epitaxy on a GaAs-substrate with (001)-orientation. The first monolayer of InAs deposited on GaAs (at 525 $^\circ\text{C}$) adopts the GaAs lattice constant. After deposition of $~1.5$ monolayers, the strain mismatch between InAs and GaAs leads to island formation \cite{Leonard1994} (Fig.\ \ref{fig:growth}(a),(b)). The islands become optically-active QDs on capping with GaAs. A two-dimensional InAs layer remains, the WL. This is the widely used SK self-assembly process. 

Here, the InAs islands are capped initially with a single monolayer of AlAs which has a higher bandgap than GaAs (Fig.\ \ref{fig:growth}(c)). Subsequently, a capping layer of $2.0\ \text{nm}$ GaAs is grown (at 500 $^\circ\text{C}$) (Fig.\ \ref{fig:growth}(d)). The additional AlAs monolayer is the only change to the standard SK protocol. For some samples, a ``flushing step" \cite{Wasilewski1999} is made following the growth of the GaAs-cap (increase of temperature to 630 $^\circ\text{C}$) (Fig.\ \ref{fig:growth}(e)). With or without flushing, the heterostructure is completed with overgrowth of GaAs (Fig.\ \ref{fig:growth}(f)).

To determine the QD-structure post growth, we carried out scanning transmission electron microscopy (STEM). Fig.\ \ref{fig:growth}(g) is a high resolution high-angle annular dark-field (HAADF) STEM-image where the contrast is related to the atomic number. The QD is the $\sim3$ nm high and $\sim30$ nm wide bright feature close to the centre of the image. The complete images with atomic resolution demonstrate that the entire structure is defect-free (see Supplementary Information). The AlAs capping layer can be clearly made out as a darker region surrounding the QD. A zoom into a region with just WL, but no QD, is shown in the right part of Fig.\ \ref{fig:growth}(g). The WL consists of InGaAs with a monolayer of AlAs contained within it. Energy dispersive X-ray spectroscopy (EDX) confirms the WL composition: In atoms are found over a $2-3$ nm thick region, yet the Al atoms are located within a $~1$ nm thick layer (Fig.\ \ref{fig:growth}(h)). These features point to highly mobile In atoms yet weakly mobile Al atoms under these growth conditions \cite{Vullum2017}. The overall thickness of the modified WL is similar to the WL of standard InGaAs QDs \cite{Blokland2009}. The In above the AlAs-layer is most likely due to In-segregation as illustrated in Fig.\ \ref{fig:growth}(c, d). The STEM-image does not indicate a transition to a Volmer-Weber growth as found in Ref.\ \onlinecite{Tsatsulnikov2000}.

We probe the electronic states initially by photoluminescence (PL) experiments. Fig.\ \ref{fig:growth}(i) shows ensemble PL from QDs grown with and without the AlAs-cap, in both cases without a flushing step. The spectra reveal the different shells (s, p, d) of the QDs. For the standard QDs, PL from the WL can be observed at $\sim925\ \text{nm}$, emission at lower wavelength is from bulk GaAs. In contrast, for the AlAs-capped QDs, the WL PL disappears. This is the first evidence for the absence of carrier confinement in the modified WL. We come to the same conclusion on flushed QDs for which the ensemble-PL is blue-shifted from $1,000-1,300$ nm to $\sim 900-980\ \text{nm}$ (Fig.\ \ref{fig:growth}(j)). Without the AlAs-capping, there is strong emission from the WL at $\sim875\ \text{nm}$. For the AlAs-capped QDs, WL emission is not observed. 

\section{PL as function of gate voltage}
\label{sec:PLVg}
The ensemble-PL measurements do not distinguish between electron and hole confinement. We make this distinction by single-QD measurements. The particular concept is to probe the QD- and WL-electron-states by gradually lowering the energy of the states with respect to the Fermi energy of a close-by Fermi sea. The QD is small enough to exhibit pronounced Coulomb blockade: electrons are added one-by-one and the QD-states are filled according to Hund's rules \cite{Warburton2000,Karrai2003}. We focus on flushed QDs, both without and with the AlAs capping layer. 

\begin{figure}
\includegraphics[width=0.9\columnwidth]{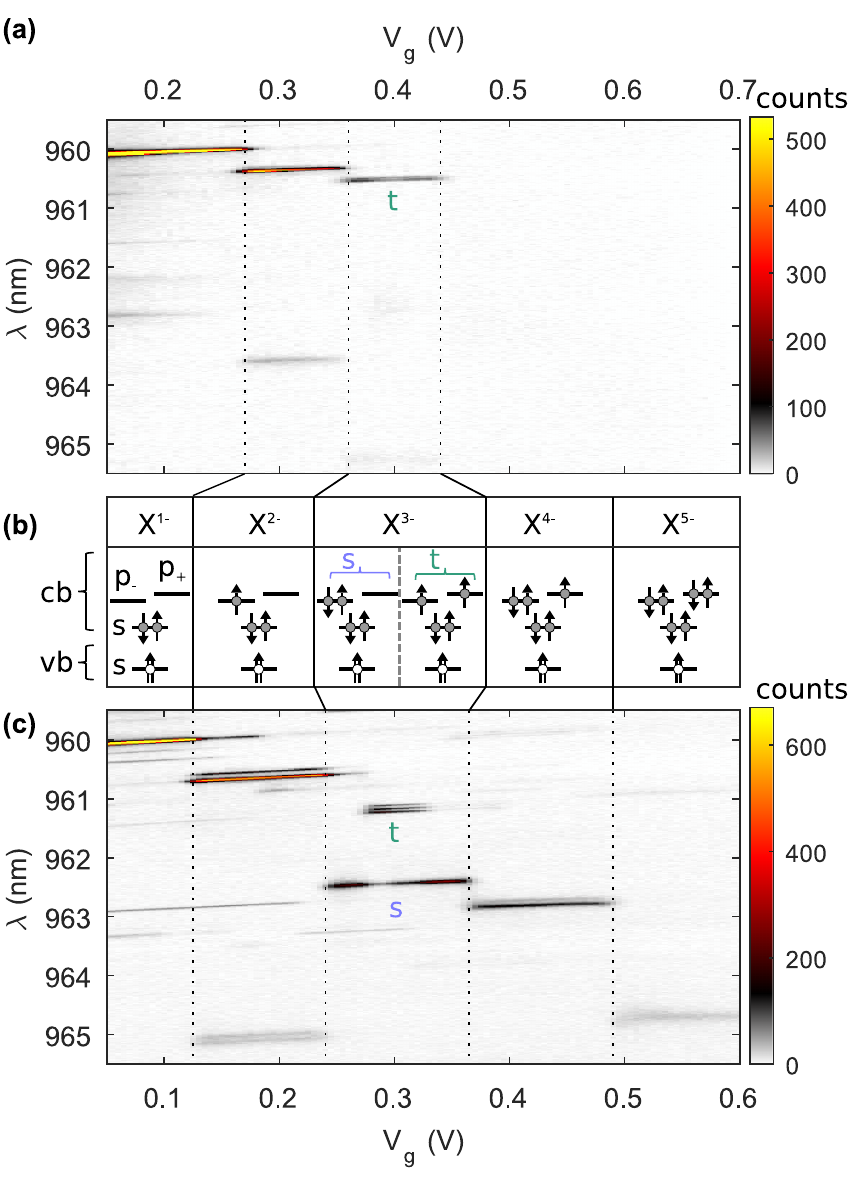}
\caption{{\bf PL as electron states are sequentially filled.} (a) PL versus gate voltage on a single, standard, flushed InGaAs QD. The plateaus correspond to Coulomb blockade \cite{Warburton2000}. (b) QD-shells and their occupation for the different exciton complexes. The triply charged exciton X$^{3-}$ has two low-lying states: a singlet (blue: s) or a triplet (green: t). (c) PL versus gate voltage on a single, AlAs-capped, flushed InGaAs QD. In the absence of WL-states for electrons, the X$^{3-}$ singlet (s) and triplet (t) as well as the highly charged exciton complexes X$^{4-}$ and X$^{5-}$ appear.}
\label{fig:PLVg}
\end{figure}

PL as function of gate voltage is shown in Fig.\ \ref{fig:PLVg} for a standard InGaAs QD (Fig.\ \ref{fig:PLVg}(a)) and for an AlAs-capped QD (Fig.\ \ref{fig:PLVg}(c)). The plateaus correspond to different charge states of the QD-exciton: in the presence of a hole, electrons fill the QD-shells sequentially. The standard QD shows charging of the neutral exciton X$^{0}$ to a net charge of $-3e$, the exciton X$^{3-}$ containing a total of four electrons and one hole. At higher gate voltage, the QD PL disappears. This is a signature that the WL becomes occupied \cite{Warburton2000,Schulhauser2004,Urbaszek2003,Rai2016,Ware2005}. 

The PL from the AlAs-capped QD is strikingly different. Charging beyond X$^{3-}$ to X$^{4-}$ and X$^{5-}$ takes place (Fig.\ \ref{fig:PLVg}(c)). The X$^{5-}$ contains a total of six electrons with fully occupied s- and p-shells. This is a novelty for QDs in this wavelength regime ($960\,\text{nm}$). Further, even the X$^{4-}$ and X$^{5-}$ result in sharp emission lines and there is no rapid loss of intensity or rapid increase in linewidth at high gate voltages. This measurement points to, first, a deep confinement potential, sufficiently deep to accommodate six electrons despite the strong Coulomb repulsions, and second, the absence of WL-states for electrons. 

At high positive bias, PL appears also at $\sim 830\ \text{nm}$ (see Supplementary Information), highly blue-shifted with respect to the QD PL, and close to the bandgap of GaAs. This PL line has a very strong Stark shift allowing us to identify it as a spatially indirect transition \cite{Kleemans2010,Rai2016} from an electron in the Fermi sea with a hole in the WL. From this line we can therefore extract the properties of the WL in the valence band. We find that the AlAs-capped QDs have a valence band WL with ionization energy 19 meV with respect to the top of the GaAs valence band (see Supplementary Information). This ionization energy is reduced with respect to the WL of standard InGaAs QDs (ionization energy $\sim 30\ \text{meV}$). The AlAs-cap eliminates any bound WL-states in the conduction band, and pushes the bound WL-states in the valence band towards the GaAs valence band edge.

\section{Triply-charged exciton}
\label{sec:Bfield}

\begin{figure*}[t]
\includegraphics[width=1.8\columnwidth]{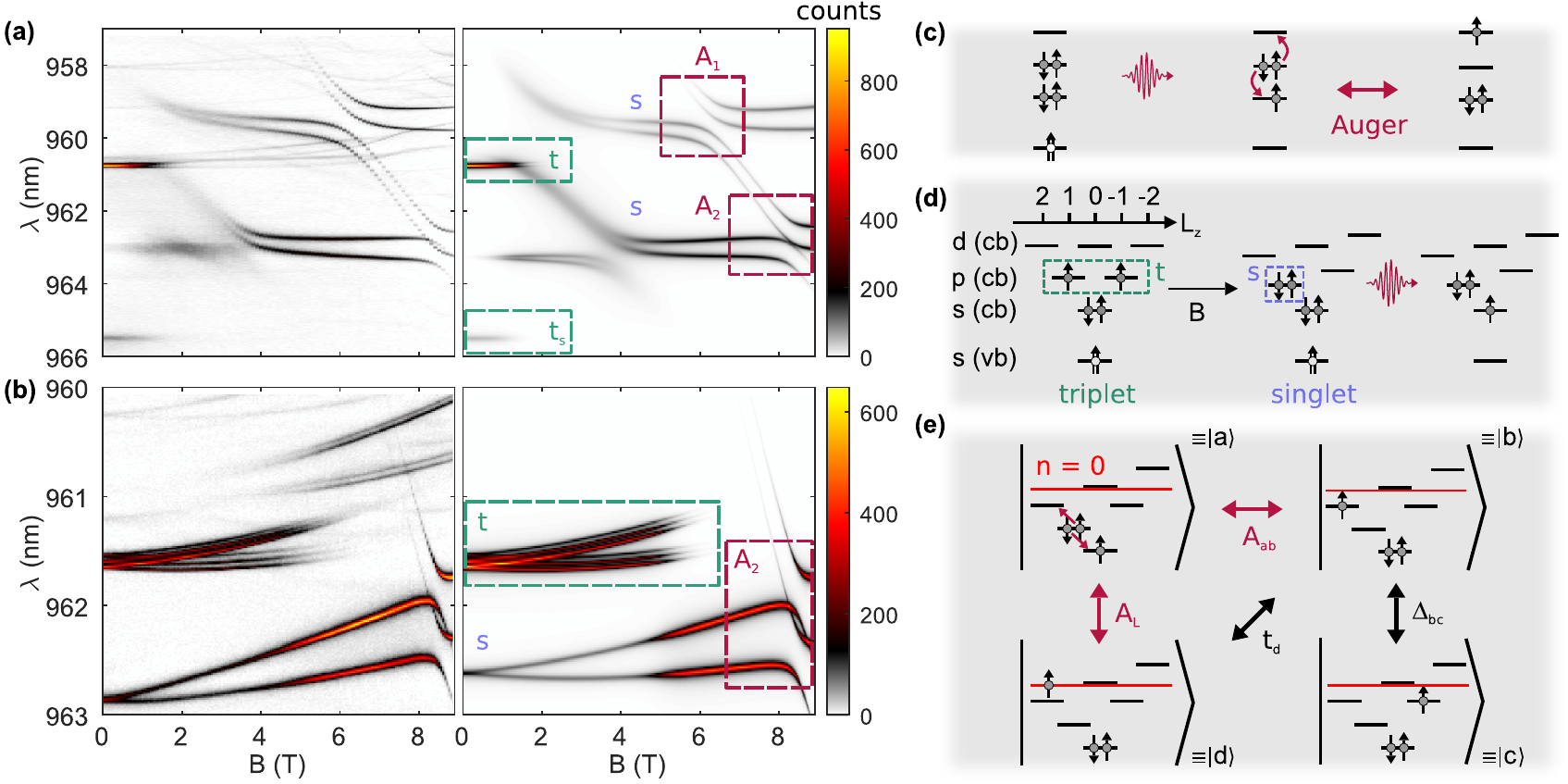}
\caption{{\bf Triply-charged exciton as probe of the QD- and QL-energy-states.} (a) Left plot: PL of the charged exciton X$^{3-}$ as a function of magnetic field for a standard InGaAs QD. Right plot: calculated magneto-PL (see Supplementary Information for parameters). (b) As (a) but for an AlAs-capped QD. Note that the line appearing at $\sim 5\ \text{T}$ and wavelength $\sim 961\ \text{nm}$ arises from $\text{X}^{2-}$, not $\text{X}^{3-}$. (c) The optical decay process of the $\text{X}^{3-}$ singlet. Following photon emission, the p-shell is doubly occupied yet there is a vacancy in the s-shell. This turns on an Auger-like coupling to a state in which a high-lying level is singly occupied (QD-shell or WL-continuum) and the s-shell is doubly occupied. In this way, the PL-process is sensitive to the high-lying state even though it is not occupied in the initial state \cite{Karrai2003}. (d) $\text{X}^{3-}$ assuming that angular momentum is a good quantum number: the p-shell has angular momentum $L_z=+1$ and $-1$; the d-shell $+2$, $0$ and $-2$. The $\text{X}^{3-}$ ground state changes from a triplet to a singlet in magnetic field. (e) The final state of the singlet X$^{3-}$. State $\ket{a}$ can couple to the d-shell of the QD via an Auger-like process (state $\ket{b}$) and to a Landau level in the WL (state $\ket{d}$). When $d_{-}$ and $p_{+}$ come into resonance, state $\ket{b}$ couples to state $\ket{c}$ where one electron occupies the $p_{+}$-sub-shell. }
\label{fig:PL_B}
\end{figure*}

We present a method to probe the high-lying energy states, for instance the QD-d-shell and WL-states, without occupying them. The method relies on an imbalance with respect to shell filling in the X$^{3-}$ final state. Following X$^{3-}$ recombination, there are two p-shell electrons yet just one s-shell electron. (Of the two s-shell electrons in the X$^{3-}$ initial state, one recombines with the hole to create a photon.) This imbalance enables Auger-like processes: one of the p-shell electrons falls into the s-shell thereby losing energy; the other p-shell electron is given exactly this energy and is promoted to a higher-lying state (Fig.\ \ref{fig:PL_B}(c)). This process will only occur if a high-lying state exists close to the right energy. If the s--p separation is $\hbar \omega_0$, the process is therefore a probe of the energy levels lying $\hbar \omega_0$ above the p-shell. Some spectroscopy is possible: the energy levels of a QD can be tuned with a magnetic field. These processes can result in large changes to the PL on charging from X$^{2-}$ to X$^{3-}$ \cite{Karrai2003,VanHattem2013}. For instance, in a QD without a d-shell, on applying a magnetic field the X$^{3-}$ PL shows a series of pronounced anti-crossings with Landau levels associated with the WL \cite{Karrai2003}: the WL is thereby probed without occupying it.

We explore initially X$^{3-}$ on standard InGaAs QDs. We measure PL in a magnetic field parallel to the growth direction (Fig.\ \ref{fig:PL_B}). For the singly and doubly charged excitons, X$^{1-}$ and X$^{2-}$, the emission splits into two lines by the Zeeman effect and blue-shifts via its diamagnetic response (see Supplementary Information). The X$^{3-}$ has a much richer strcuture (Fig.\ \ref{fig:PL_B}(a)). At zero magnetic field, the X$^{3-}$ has a configuration with two electrons in the QD-s-shell and two electrons in the p-shell. According to Hund's rules, the ground state electrons occupy different p-sub-shells with parallel spins (a spin-triplet) and two emission lines result, split by the large electron-electron exchange energy, denoted as $t$ (triplet) and $t_s$ (triplet satellite) in Fig.\ \ref{fig:PL_B}(a) \cite{Karrai2003}. On increasing the magnetic field, the degeneracy (or near degeneracy) of the p-sub-shells is lifted. In the Fock-Darwin model \cite{Miller1997,Kouwenhoven2001,Karrai2003}, the $p_{-}$-sub-shell (angular momentum $L_z=+1$) moves down in energy by $-\frac{1}{2}\hbar\omega_c$ while the $p_{+}$-sub-shell (angular momentum $L_z=-1$) moves up in energy by $+\frac{1}{2}\hbar\omega_c$ (Fig.\ \ref{fig:PL_B}(d)). Here $\hbar \omega_c$ is the electron cyclotron energy. Once this splitting becomes large enough the X$^{3-}$ ground state turns from a triplet to a singlet where two electrons of opposite spin populate the lower p-sub-shell (Fig.\ \ref{fig:PL_B}(d)) \cite{Karrai2003}. The transition from triplet to singlet ground state occurs at $\sim1.3\ \text{T}$ (Fig.\ \ref{fig:PL_B}(a)). The singlet (and not the triplet) ground state represents the probe of the higher lying electronic states.

The magnetic field dependence of the X$^{3-}$ singlet-PL-spectrum on a standard InGaAs QD shows several anti-crossings (Fig.\ \ref{fig:PL_B}(a)). We develop a model to describe the X$^{3-}$ final state including Coulomb interactions within a harmonic confinement and couplings to a WL-continuum (see Supplementary Information). In addition to the energies of the transitions, the linewidths are a powerful diagnostic. The spectrally narrow PL-lines arise from intra-QD-processes; the spectrally broad PL-lines from QD--WL-continuum coupling as the continuum of WL-states facilitates rapid dephasing \cite{Karrai2003,Urbaszek2004}. 

The singlet emission at $\sim1.3\ \text{T}$ is spectrally broad: this signifies that the final state couples to the WL-continuum. There is an anti-crossing at $\sim3\ \text{T}$ with a state with a linear magnetic field dispersion. This signifies a hybridization with the 0th WL-Landau-level (Fig.\ \ref{fig:PL_B}(e)). A second singlet emission line appears at higher energy, and there are two further anti-crossings at high magnetic field ($A_1$ and $A_2$ in Fig.\ \ref{fig:PL_B}(a)). We exclude that these processes are caused by a hybridization with the WL since the optical emission stays narrow in this regime. The first part of the explanation is an Auger-like process within the QD itself (Fig.\ \ref{fig:PL_B}(e)). The optical decay of the X$^{3-}$ singlet leaves behind two electrons in the lower p-sub-shell and one electron in the s-shell (state $\ket{a}$). This final state can couple to state $\ket{b}$ via an Auger-like process where one p-electron fills the vacancy in the s-shell and the other goes up into the d-shell. This coherent coupling between the two basis states $\ket{a}$ and $\ket{b}$ leads to two eigenstates after optical decay and thus explains the second emission line at higher energy. The second part of the explanation involves the single particle states. With increasing magnetic field, the $d_{-}$-sub-shell of the QD moves down in energy with a dispersion of $-\hbar\omega_c$ while the $p_{+}$-sub-shell moves up with $\frac{1}{2}\hbar\omega_c$. In the Fock-Darwin model, angular momentum is a good quantum number and $d_{-}$ and $p_{+}$ therefore cross. Experimentally, this is not the case: there is a small anti-crossing. This is not surprising for a real QD where there is no exact rotational symmetry. To describe this, we introduce basis state $\ket{c}$ (with an electron in the $p_{+}$- rather than the $d_{-}$-shell) and a small coupling between states $\ket{b}$ and $\ket{c}$ to describe the symmetry breaking. This leads to the two characteristic anti-crossings ($A_1$, $A_2$) of the singlet emission pair with a line with dispersion of approximately $-\frac{3}{2}\hbar\omega_c$. 

An analytic Hamiltonian describing all these processes is given in the Supplementary Information. Using realistic parameters for the QD, the model (Fig.\ \ref{fig:PL_B}(a)) reproduces the X$^{3-}$ PL extremely well. This strong agreement allows us to extract the key QD parameters from this experiment: the electron s--p splitting ($\hbar\omega_0=24.1$ meV) and the electron effective mass ($0.07 m_{o}$). We are also to conclude that the potential is sub-harmonic: the p--d splitting is smaller than the s--p splitting.

With this understanding of the X$^{3-}$, we turn to the spectra from an AlAs-capped QD (Fig.\ \ref{fig:PL_B}(b)). As for the standard InGaAs QD, there is a transition from triplet to singlet X$^{3-}$ ground state, albeit at higher magnetic fields. In complete contrast to the standard InGaAs QD, the hybridization with a Landau level is not observed. This is powerful evidence that the electron WL-states no longer exist. 

The X$^{3-}$ from the AlAs-capped QD is revealing in a number of other respects. First, the X$^{3-}$ singlet state shows one Zeeman-split line, not two as for the standard InGaAs QD. This is evidence that the $\ket{a}-\ket{b}$ coupling is suppressed on account of the energies: state $\ket{b}$ lies at too high an energy to couple to state $\ket{a}$ (see Supplementary Information). A large ratio between $\ket{b} - \ket{a}$ energy splitting and coupling strength leads to a very weak emission from the second line, strongly red-shifted for a positive $\ket{b} - \ket{a}$ energy splitting. The absence of a second singlet emission line is evidence that the p--d splitting is larger than the s--p splitting, a super-harmonic potential. This is consistent with the thin, AlAs-layer in the STEM-characterization (Fig.\ \ref{fig:growth}(g)) which bolsters the lateral confinement; and also the ensemble-PL where the AlAs-cap blue-shifts the d-shell more than the p-shell (Fig.\ \ref{fig:growth}(i)). Second, the X$^{3-}$ singlet and triplet X$^{3-}$-PL-lines appear simultaneously at low magnetic field (Fig.\ \ref{fig:PL_B}(b)) yet there is a rather abrupt transition for the standard InGaAs QD (Fig.\ \ref{fig:PL_B}(a)). This is an indication that relaxation to the exciton ground state is slower for the AlAs-capped QDs. This may also be related to the WL: electrons in the WL can mediate spin relaxation and without the WL, this process is turned off. Finally, the X$^{3-}$ exciton in the AlAs-capped QD has a very pronounced fine structure splitting: the splitting of the X$^{3-}$ triplet into three lines is a prominent feature (Fig.\ \ref{fig:PL_B}(b)). This particular fine structure originates from electron-hole exchange in the initial exciton state \cite{Urbaszek2003}, and its increase relative to standard InGaAs QDs is indicative of a stronger electron-hole confinement \cite{Bayer2002,Franceschetti1998}. 

We model X$^{3-}$ in the AlAs-capped QD with the model developed for the standard InGaAs QD. The coupling to the Landau level is set to zero. A small perturbation is included to account for the anharmonicity of the confinement potential. The model describes the experimental results extremely well (Fig.\ \ref{fig:PL_B}(b)). The model determines the electron s$-$p splitting as $\hbar\omega_0=27.5$ meV.

\section{Temperature dependence}
\label{sec:Tdep}
\begin{figure}
\includegraphics[width=0.9\columnwidth]{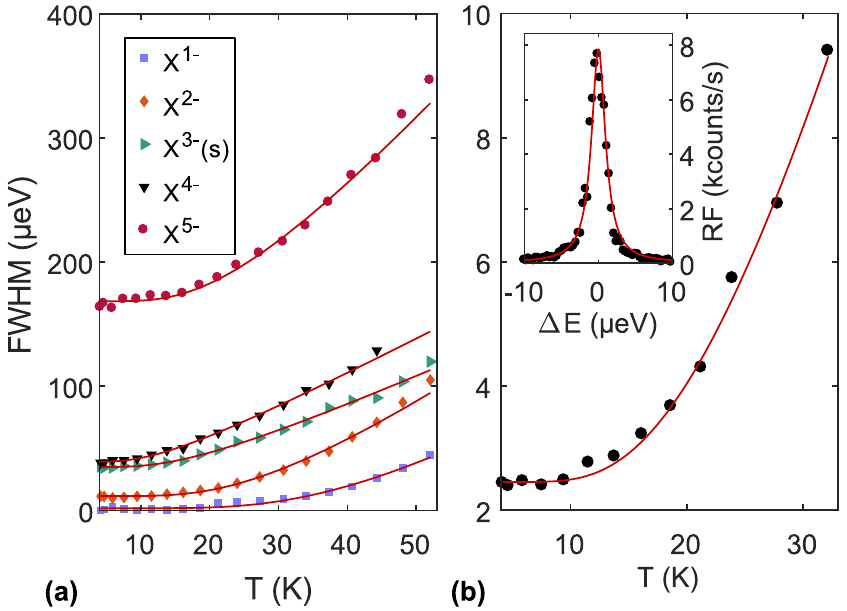}
\caption{{\bf Linewidths of excitons in an AlAs-capped QD.} (a) PL-linewidth (FWHM) of the charged excitons as function of temperature. The red lines represent a fit to a model that describes the interaction with acoustic phonons (see Supplementary Information). (b) Linewidth (FWHM) of the singly charged exciton X$^{1-}$ in resonance fluorescence as a function of temperature. The insert shows an example resonance fluorescence measurement at $4.2\ \text{K}$ with a linewidth of $2.3\ \mu\text{eV}$. This measurement was carried out at low excitation power (coherent scattering regime). The saturation count rate obtained under resonant excitation is $60$ kcounts/s.}
\label{fig:Tdep}
\end{figure}
The temperature dependence of the exciton linewidths is a further probe of the coupling to continuum states. Linewidths of excitons in standard InGaAs QDs strongly increase with temperature as soon as hybridization with a WL is present \cite{Urbaszek2004}. Such a temperature broadening was observed for exciton complexes even with modest charge \cite{Urbaszek2004}, for instance X$^{2-}$. For an AlAs-capped QD, we measure the PL-linewidth of all charged excitons (X$^{1-}$ to X$^{5-}$) as a function of temperature (Fig.\ \ref{fig:Tdep}(a)). Even for the highly charged excitons X$^{4-}$ and X$^{5-}$, the temperature-induced broadening is much weaker than that for charged excitons beyond X$^{1-}$ in standard InGaAs QDs which show a strong, linear temperature dependence \cite{Urbaszek2004}. Instead, the linewidths are described well by a model which considers a localized exciton and dephasing via acoustic phonon scattering \cite{Rudin2006}. This too is evidence that the WL-states for electrons no longer exist.

Finally, we measure the linewidth of the singly charged exciton (X$^{1-}$) with resonant excitation, detecting the resonance fluorescence \cite{Kuhlmann2013,Kuhlmann2013a} (Fig.\ \ref{fig:Tdep}(b)). The resonance fluorescence linewidth increases with temperature above $\sim 10$ K, indicative of acoustic phonon scattering. At $4.2$ K, the linewidth ($2.3\ \mu$eV) is similar to the linewidth of the very best InGaAs QDs \cite{Kuhlmann2013}. This shows that the AlAs-capped QDs retain the very low charge noise achieved for standard InGaAs QDs \cite{Kuhlmann2013,Kuhlmann2014}. This is important: the AlAs-capped QDs have slow exciton dephasing and weak spectral fluctuations such that they are completely compatible with applications which place stringent requirements on the quality of the single photons.

\section{acknowledgements}
\label{sec:acknowledgement}
MCL, IS and RJW acknowledge financial support from NCCR QSIT and from SNF Project No.\ 200020\_156637. This project received funding from the European Union Horizon 2020 Research and Innovation Programme under the Marie Sk\l{}odowska-Curie grant agreement No.\ 747866 (EPPIC). SS, JR, AL, and ADW gratefully acknowledge financial support from the grants DFH/UFA CDFA05-06, DFG TRR160, DFG project 383065199, and BMBF Q.Link.X 16KIS0867.
\vspace{1em}
\section{author contributions}
\label{sec:constrib}
SS and AL initiated the project. SS, JR and AL carried out the QD-growth and ensemble-PL-measurements together with ADW. TD, AK and BEK carried out the STEM and EDX analysis. MCL and IS carried out the single-QD-spectroscopy. MCL, IS, AL and RJW developed concepts and established the links between the various aspects of the project. MCL and RJW wrote the paper with input from all authors.

\bibliography{lit_no_WL}

\onecolumngrid
\newpage

\makeatletter 
\renewcommand{\thefigure}{S\@arabic\c@figure}
\makeatother

\makeatletter 
\renewcommand{\thetable}{S\@arabic\c@table}
\makeatother

\makeatletter 
\renewcommand{\theequation}{S\@arabic\c@equation}
\makeatother

\section{Supplementary Information}
\subsection{Sample Growth, Fabrication, and STEM-measurements}
\label{sec:growth_details}
The sample heterostructure is grown with molecular beam epitaxy (MBE) on a GaAs wafer with (001)-orientation. The overall growth conditions are similar to those described in Ref.\ \onlinecite{Ludwig2018}.
\begin{table}[b]
\begin{ruledtabular}
\begin{tabular}{lccc} 
Material & Thickness (nm) & Temperature ($^\circ\text{C}$) & Rate (\AA)\\\cline{1-4} 
GaAs & 50 & 600 & 2.0\\
AlAs/GaAs & 30$\times$(2/2) & 600 & 1.0/2.0\\
GaAs & 50 & 600 & 2.0\\
n-GaAs & 300 & 600 & 2.0\\
GaAs & 5 & 575 & 2.0\\
GaAs & 25 & 600 & 2.0\\
InAs QDs & -- & 525 & --\\
AlAs-capping & 0.3 & 525 & 1.0\\
GaAs-capping & 2 & 500 & 2.0\\
flushing step & -- & 600 & --\\
GaAs & 8 & 600 & 2.0\\
AlAs/GaAs & 30$\times$(3/1) & 600 & 1.0/2.0\\
GaAs & 10 & 600 & 2.0\\
\end{tabular}
\caption{\label{table_structure}Description of the sample. The different layers of the heterostructure are listed in the order of the growth.}
\end{ruledtabular}
\end{table}
The heterostructure, together with growth temperatures and growth rates of individual layers, is given in Table \ref{table_structure}. To flatten the wafer surface, an AlAs/GaAs-superlattice is grown first. A spacer of $50\ \text{nm}$ GaAs separates the superlattice from a silicon-doped (n-type) back-gate ($300\ \text{nm}$). QDs are separated from the back-gate by a tunnel barrier of $30\ \text{nm}$ GaAs. The QDs (growth described in the main text) are overgrown with an additional $8\ \text{nm}$ of GaAs. To keep the current through the diode structure low, a tunnel barrier (another AlAs/GaAs-superlattice) is grown above the QDs. The heterostructure is completed with a thin GaAs capping layer. 

For electric charge control of the QDs, a semi-transparent Schottky top-gate ($\sim6\ \text{nm}$ Au) is deposited on part of the sample. An ohmic contact to the back-gate is fabricated by annealing In-solder for $5\ \text{min}$ at $370\ ^\circ\text{C}$ in a forming gas atmosphere.

\begin{figure*}[t]
\includegraphics[width=0.9\columnwidth]{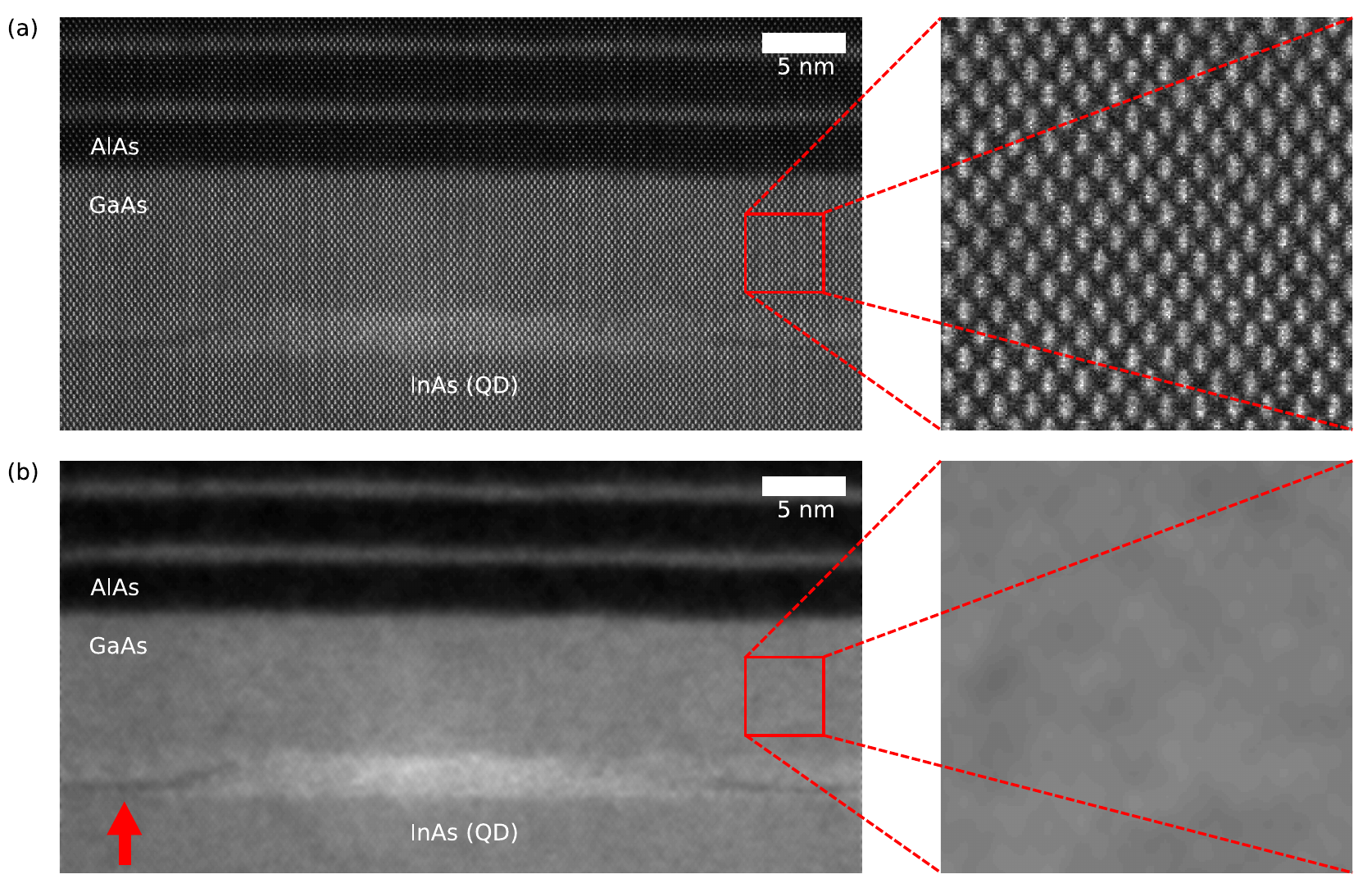}
\caption{(a) Atomic resolution HAADF STEM image of an AlAs-capped quantum dot (QD) grown with a flushing step. The QD has a high indium concentration and thus appears bright in the STEM-image. Above the QD, an AlAs/GaAs-superlattice is visible. (b) A Gaussian smoothing of the STEM-image (6 pixel standard deviation) enhances the contrast in the STEM-image. At the edges of the QD, the AlAs capping layer appears as a darker region in the STEM-image (indicated by the red arrow).}
\label{fig:TEMsupp}
\centering
\end{figure*}

For scanning transmission electron microscopy (STEM) investigations, an electron-transparent lamella was prepared by conventional mechanical polishing followed by argon ion milling. The high resolution HAADF STEM-image was acquired using a FEI Titan G2 equipped with a Schottky field emission gun operated at 200 kV, a Cs probe corrector (CEOS DCOR). The annular dark-field detector semi-angle used was 69.1 mrad. For STEM, a series of 10 images were recorded with a short acquisition time, aligned and summed using the Velox software (Thermo Scientific) to improve the quality of the image. In the STEM-HAADF image, the intensity of the atomic columns is approximately proportional to the square of the atomic number \cite{Pennycook1999}. STEM measurements are combined with an energy dispersive X-ray (EDX) measurement (same microscope). Fig.\ \ref{fig:TEMsupp} shows a STEM-image of an AlAs-capped QD. The image has atomic resolution showing that QD and its surroundings are defect-free. For better visibility, we increase the contrast of the original STEM-image by Gaussian smoothing. The result corresponds to a STEM-image taken with lower spatial resolution and is shown in Fig.\ \ref{fig:TEMsupp}(b). In the STEM-images, a single QD is visible as a bright (In-rich) region. Due to the lower atomic number of aluminium, the AlAs-capping surrounding the QD appears darker. Part of the AlAs/GaAs-superlattice grown above the QDs is visible as an alternating sequence of bight and dark regions in the STEM-image.

\subsection{Modeling of the Magneto-PL-measurements}
\label{sec:supp_MagnetoPL}
We present the calculation which describes the magneto-PL of the X$^{3-}$ exciton complex. The experimental data are shown in Fig.\ 3 of the main text. The energy of the optical emission lines is given by the energy difference between the initial exciton state before optical recombination and the energy of the electron configuration in the final state after optical recombination. We calculate the energies of the initial and final states separately.

\begin{figure}[b]
\includegraphics[width=0.9\columnwidth]{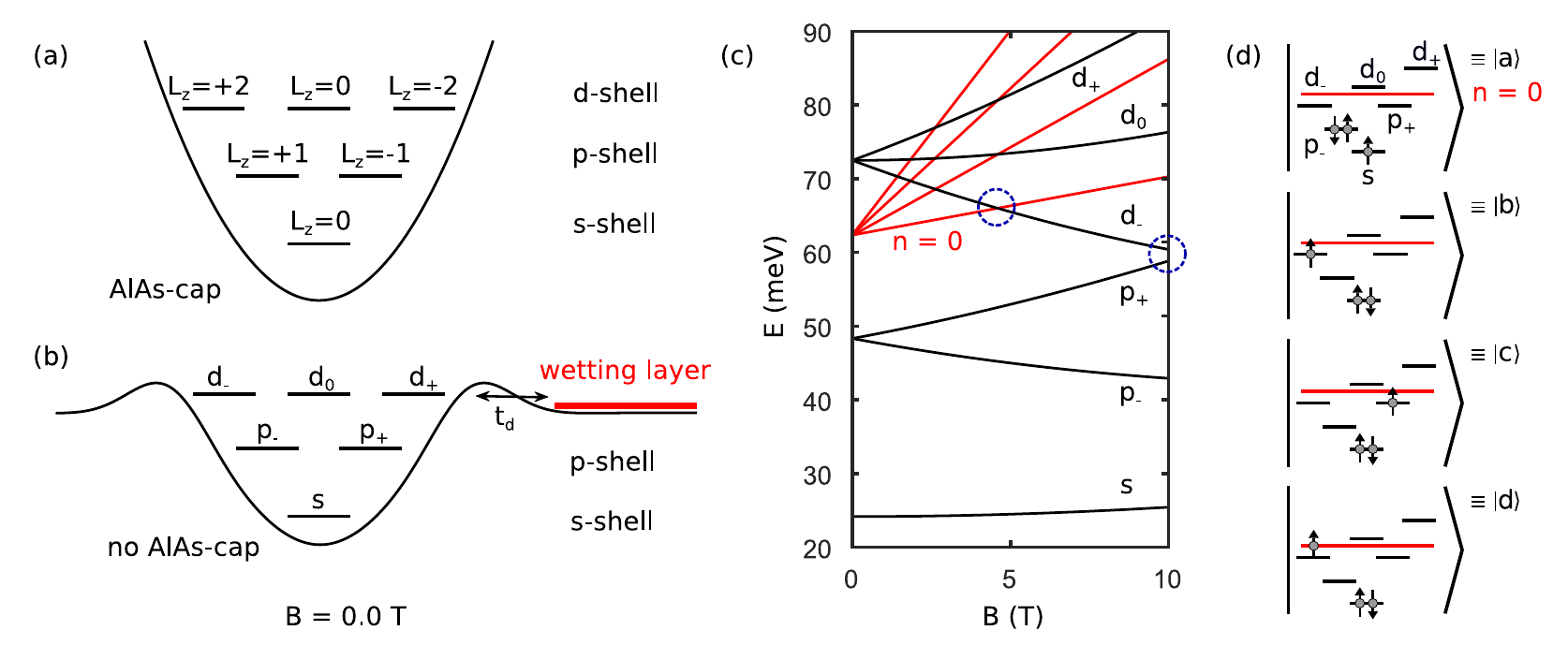}
\caption{(a) Schematic conduction levels for an AlAs-capped QD. The different shells (s, p, d) are labeled with the corresponding angular momentum quantum number $L_z$. (b) Schematic conduction levels for a standard QD with a WL. States of the WL-continuum are indicated in red. (c) Black curves: single particle dispersion for a symmetric two-dimensional harmonic oscillator ($\hbar\omega_0\approx24\ \text{meV}$) as a function of magnetic field (Fock-Darwin spectrum). Red curves: Dispersion of the WL-Landau-levels. $n=0$ denotes the lowest Landau level. (d) The four basis states $\ket{a}$, $\ket{b}$, $\ket{c}$, and $\ket{d}$ used to describe the possible final states after X$^{3-}$ recombination. The basis states are shown at high magnetic field.}
\label{fig:suppl_intro}
\centering
\end{figure}

The focus is the X$^{3-}$ singlet exciton. On recombination of an s-shell electron with an s-shell hole, there is a vacancy in the electron s-shell; the p-shell is doubly occupied. This special situation allows Auger-like processes to take place: one of the p-shell electrons falls into the vacancy, the other is promoted to a higher level. These processes admix the QD-d-shell and, should they exist, the WL-Landau-levels, into the available final states. The X$^{3-}$ final states have therefore a rich structure. As we argue in the main text, they provide an ideal way of exploring the single particle states at energies well above both the s- and p-shells. 

Our approach is to focus initially on the behaviour of X$^{3-}$ in a QD with a WL. In this case, there is both a confined d-shell and a WL which quantises into Landau levels in a magnetic field. This is the most complex situation. We develop a model to describe this situation by considering hybridizations between different final states after optical recombination. This allows us to uncover precisely how the X$^{3-}$ spectrum depends on the d-shell and on the WL-Landau-levels: we develop understanding of our probe of the high-energy states. We then apply this model to a QD in a sample with a modified WL. The absence of features related to the WL enables us to demonstrate that the WL as far as electron states are concerned is no longer present.

\subsubsection{The X$^{3-}$ final states}
Assuming that the QD has a lateral parabolic confinement potential of the form $V(r)=\frac{1}{2}m_e^*\omega_0^2 r^2$, the single particle energies of the different QD shells follow the well known Fock-Darwin spectrum \citep{Fock1928,Darwin1930,Kouwenhoven2001}:
\begin{equation}
\label{Eq_FockD}
E_{n,L}=\left(2n+\abs{L_z}+1\right)\hbar\omega_1-\frac{1}{2}L_z\hbar\omega_c.
\end{equation}
Here, $n$ is the radial quantum number and $L_z$ the angular momentum quantum number, $\omega_1=\sqrt{\omega_0^2+\left(\frac{\omega_c}{2}\right)^2}$, where $\hbar \omega_c=\hbar eB/m_{e}^{*}$ is the electron cyclotron energy, $B$ the magnetic field, and $m_{e}^{*}$ the electron effective mass. In Fig.\ \ref{fig:suppl_intro}(c) the magnetic field dispersion described by Eq.\ \ref{Eq_FockD} is shown for different QD shells. The corresponding wave functions are given by \citep{Pfannkuche1993,Tannoudji1977,Kouwenhoven2001}
\begin{equation}
\label{Eq_2D_Psi}
\ket{\Psi_{n,L_z}\left(r,\phi\right)}=\frac{e^{iL_z\phi}}{\sqrt{\pi}L_e}\sqrt{\frac{n!}{\left(n+\abs{L_z}\right)!}}e^{-\frac{r^2}{2L_e^2}}\left(\frac{r^2}{L_e^2}\right)^{\abs{L_z}/2}L_n^{\abs{L_z}}\left(\frac{r^2}{L_e^2}\right).
\end{equation}
The terms $L_n^{\abs{L_z}}$ denote generalised Laguerre polynomials. The parameter $L_{e}=\sqrt{\hbar/(\omega_1m_{e}^*)}$ is the effective length of the wave function.

We consider initially three basis configurations $\ket{a}$, $\ket{b}$, and $\ket{c}$ to describe the final state of the X$^{3-}$ (singlet) exciton. These states and a schematic shell structure of a QD are illustrated in Fig.\ \ref{fig:suppl_intro}. We calculate the energies of states $\ket{a}$, $\ket{b}$, and $\ket{c}$ as the sum of their corresponding single particle energies and the energies caused by Coulomb interactions between different electrons.

Basis state $\ket{a}$ has one electron in the s-shell and two electrons in the p$_-$-shell (angular momentum $L_z=+1$). According to Eq.\ \ref{Eq_FockD}, the sum of all single particle energies is given by $E_{a}^{0}=5\hbar\omega_1-\hbar\omega_c$. State $\ket{b}$ has two electrons in the s-shell and one electron in the d$_-$-shell (angular momentum $L_z=+2$). Its single particle energy is the same as that of state $\ket{a}$, i.e.\ $E_{b}^{0}=E_{a}^{0}$. State $\ket{c}$ has two electrons in the s-shell and one electron in the p$_+$-shell (angular momentum $L_z=-1$). Its single particle energy is given by $E_{c}^{0}=4\hbar\omega_1+\frac{1}{2}\hbar\omega_c$. The single particle energy of state $\ket{c}$ increases in a magnetic field with $+\frac{1}{2}\hbar\omega_c$ whereas that of states $\ket{a}$ and $\ket{b}$ decreases with $-\hbar\omega_c$. 

The multi-particle nature of the QD-states gives rise to carrier-carrier Coulomb interactions within the QD, both for the initial exciton state (four electrons and one hole in the QD) and the final states $\ket{a}$, $\ket{b}$, and $\ket{c}$ (three electrons in the QD). We calculate the corresponding energy corrections to the single particle energies in first order perturbation theory \citep{Warburton1998,Hawrylak2003}. These energies are the direct carrier-carrier Coulomb interaction given by the relations
\begin{equation}
\label{Eq_direct_Coulomb}
E_{ij}^d = \frac{e^2}{4\pi\epsilon_0\epsilon_r}\int\int \frac{\abs{\Psi_i\left(\textbf{r}_1\right)}^2\cdot\abs{\Psi_j\left(\textbf{r}_2\right)}^2}{\abs{\textbf{r}_1-\textbf{r}_2}}\ d\textbf{r}_1\ d\textbf{r}_2
\end{equation}
and the exchange interaction given by
\begin{equation}
\label{Eq_exchange_Coulomb}
E_{ij}^x = \frac{e^2}{4\pi\epsilon_0\epsilon_r}\int\int \frac{\Psi_i\left(\textbf{r}_1\right)^*\Psi_j\left(\textbf{r}_2\right)^*\Psi_i\left(\textbf{r}_2\right)\Psi_j\left(\textbf{r}_1\right)}{\abs{\textbf{r}_1-\textbf{r}_2}}\ d\textbf{r}_1\ d\textbf{r}_2.
\end{equation}
For the calculation we use single particle wave functions for a symmetric parabolic confinement potential (see Eq.\ \ref{Eq_2D_Psi}) and construct fully anti-symmetrized wave functions via Slater determinants. In case of state $\ket{a}$, for instance, this yields:
\begin{align}\label{Eq_state_a}
\ket{a}=\frac{1}{\sqrt{6}}\begin{vmatrix}
s_1\uparrow_1 & s_2\uparrow_2 & s_3\uparrow_3\\
p_1\uparrow_1 & p_2\uparrow_2 & p_3\uparrow_3\\
p_1\downarrow_1 & p_2\downarrow_2 & p_3\downarrow_3
\end{vmatrix}= 
\begin{matrix}
\frac{1}{\sqrt{6}}[&\uparrow_1\uparrow_2\downarrow_3\left(s_1p_2p_3-p_1s_2p_3\right)\\
+&\uparrow_1\downarrow_2\uparrow_3\left(p_1p_2s_3-s_1p_2p_3\right)\\
+&\downarrow_1\uparrow_2\uparrow_3\left(p_1s_2p_3-p_1p_2s_3\right)\;].\\
\end{matrix}
\end{align}
Here, the indices label the different particles, the arrows ($\uparrow$, $\downarrow$) represent the electron spin, and the orbital wave function is represented by the shell label $s$, $p$, and $d$ (see Fig.\ \ref{fig:suppl_intro}). For the basis states $\ket{a}$, $\ket{b}$, and $\ket{c}$ the sum of all the direct and the exchange energies is given by:
\begin{align}
\label{Eq_Ec}
&E_{a}^C=\frac{31}{16}\frac{e^2}{4\pi\epsilon_0\epsilon_r}\frac{1}{L_e}\sqrt{\frac{\pi}{2}},\\
&E_{b}^C=\frac{67}{32}\frac{e^2}{4\pi\epsilon_0\epsilon_r}\frac{1}{L_e}\sqrt{\frac{\pi}{2}},\\
&E_{c}^C=\frac{9}{4}\frac{e^2}{4\pi\epsilon_0\epsilon_r}\frac{1}{L_e}\sqrt{\frac{\pi}{2}}.
\end{align}
In these expressions, $e$ is the elementary charge, $\epsilon_0$ the permittivity of vacuum, and $\epsilon_r$ the relative permittivity. 

We focus now on the couplings between the basis states $\ket{a}$, $\ket{b}$, and $\ket{c}$. As illustrated in the main text, states $\ket{a}$ and $\ket{b}$ are coupled by an Auger-like process. For state $\ket{a}$, two electrons occupy the p$_-$-shell (with $L_z=+1$). In the Auger-like process, one of these two electrons goes down to the s-shell (with $L_z=0$) while the other one goes up to the d$_-$-shell (with $L_z=+2$). This process conserves angular momentum and leads to a coupling between $\ket{a}$ and $\ket{b}$. The corresponding matrix element is given by:
\begin{equation}
\label{Eq_Auger_ab}
A_{ab}=\braket{p_1^{L_z=+1}p_2^{L_z=+1}\mid\hat{H}_{c}\mid s_1^{L_z=0}d_2^{L_z=+2}} =  -\frac{5\sqrt{2}}{32}\frac{e^2}{4\pi\epsilon_0\epsilon_r}\frac{1}{L_e}\sqrt{\frac{\pi}{2}},
\end{equation}
where s, p and d label the single particle shell with particle number in the subscript and the angular momentum quantum number in the superscript. $\hat{H}_{c}$ is the two-particle Coulomb operator given by:
\begin{align}
\label{Eq_HC}
\hat{H}_{c}&=\frac{1}{2}\sum_{i}\sum_{j\neq i}\hat{H}_{ij}, \text{with}\\
\hat{H}_{ij}&=\frac{e^2}{4\pi\epsilon_0\epsilon_r}\frac{1}{\abs{\textbf{r}_i-\textbf{r}_j}},
\end{align}
where $\textbf{r}_i$ are the coordinates of the interacting particles. 

\begin{figure}[t]
\includegraphics[width=1.0588\columnwidth]{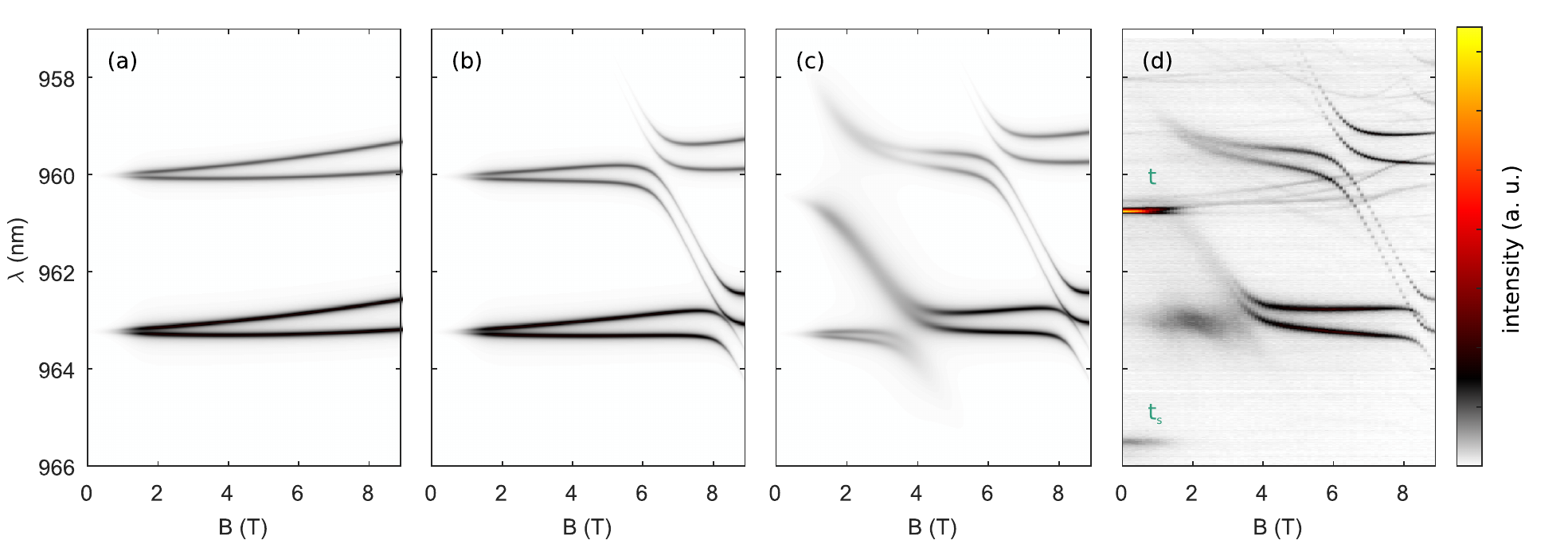}
\caption{Description of X$^{3-}$ in a QD with WL. (a) Simulation of the X$^{3-}$ emission spectrum without the coupling term $\Delta_{bc}$ between states $\ket{b}$ and $\ket{c}$ and without couplings between QD- and WL-states ($A_L$, $t_d$). (b) Simulation  including the coupling term $\Delta_{bc}$, yet without the couplings between QD- and WL-states. (c) Final simulation of the X$^{3-}$ emission spectrum including all coupling terms. (d) Measured X$^{3-}$ emission spectrum as a function of magnetic field. The simulation shown in (a-c) focuses on the X$^{3-}$ singlet, emission of the X$^{3-}$ triplet at low magnetic field ($t$, $t_s$) is not included.}
\label{fig:simulation_explain}
\end{figure}

Fig.\ \ref{fig:simulation_explain}(a) shows a numerical simulation based on the model developed so far: a coupling between $\ket{a}$ and $\ket{b}$; no coupling of states $\ket{a}$ and $\ket{b}$ to state $\ket{c}$. (State $\ket{c}$ is therefore irrelevant at this point in the calculation.) Without the $\ket{a} \leftrightarrow \ket{b}$ coupling, there is just a single pair of emission lines. (The splitting within the pair into two lines arises from the spin Zeeman effect.) With the $\ket{a} \leftrightarrow \ket{b}$ coupling, there is no longer just one pair of emission lines but two, Fig.\ \ref{fig:simulation_explain}(a). This feature describes part of the experimental data, Fig.\ \ref{fig:simulation_explain}(d). This demonstrates both that the d-shell exists and that the Auger-like process admixes the d-shell into the X$^{3-}$ final states.

The experimental X$^{3-}$ emission from a standard QD with WL (Fig.\ \ref{fig:simulation_explain}(d)) shows a richer structure than that described with just $\ket{a} \leftrightarrow \ket{b}$ coupling (Fig.\ \ref{fig:simulation_explain}(a)). In the experiment, there are several anti-crossings at high magnetic field along with a complex structure at low magnetic field. This leads us to the conclusion that additional couplings must be introduced to describe the experimental results. 

First, we consider the coupling between states $\ket{b}$ and $\ket{c}$. For a perfectly symmetric harmonic confinement potential this coupling is zero since states $\ket{b}$ and $\ket{c}$ have different angular momenta. Nevertheless, the experiment points to a coupling in the present case. This coupling represents a slight asymmetry in the confinement potential of the QD, since in this case angular momentum is not a good quantum number \citep{Tannoudji1977}. In our simulations we assume a constant coupling term $\Delta_{bc}$ between the states $\ket{b}$ and $\ket{c}$. A numerical simulation taking the coupling $\Delta_{bc}$ into account is shown in Fig.\ \ref{fig:simulation_explain}(b). The coupling accounts for the pronounced anti-crossings ($A_1$, $A_2$ in the main text) in the X$^{3-}$ emission lines at high magnetic field.

We note that we neglect any direct coupling between states $\ket{a}$ and $\ket{c}$. At low magnetic field, $\ket{c}$ is energetically far away from $\ket{a}$. At high magnetic field, when $\ket{c}$ comes energetically close to the states $\ket{a}$ and $\ket{b}$, a coupling between $\ket{a}$ and $\ket{c}$ would be both a two particle and angular momentum non-conserving process. On this basis, we assume that a $\ket{a} \leftrightarrow \ket{c}$ coupling is much weaker than the $\ket{b} \leftrightarrow \ket{c}$ coupling. (The $\ket{b} \leftrightarrow \ket{c}$ coupling is a single particle process and is important when the energies of the $d_-$-shell and the $p_+$-shell come into resonance, Fig.\ 3(e) of the main text and Fig.\ \ref{fig:suppl_intro}(c).) Of course, via the $\ket{a} \leftrightarrow \ket{b}$ and $\ket{b} \leftrightarrow \ket{c}$ couplings, states $\ket{a}$ and $\ket{c}$ will anti-cross. 

The $\ket{b} \leftrightarrow \ket{c}$ coupling does not account for the behaviour at magnetic fields of $1-4$ T as revealed by a comparison of the calculation (Fig.\ \ref{fig:simulation_explain}(b)) with the experiment (Fig.\ \ref{fig:simulation_explain}(d)). We cannot account for this behaviour within the $\ket{a}$, $\ket{b}$, $\ket{c}$ basis. The experiment shows features with a strong negative energy dispersion with increasing magnetic field. These are suggestive of Landau levels in the final state \cite{Karrai2003}. Landau levels are features of a quantised two-dimensional continuum: they are associated with the WL. 

Next, we consider the coupling between WL-Landau-levels and the QD. An Auger-like coupling between state $\ket{a}$ and WL-Landau-level \cite{Karrai2003} is illustrated in Fig.\ 3(e) of the main text. The Landau levels of the two-dimensional WL have a dispersion given by $\hbar\omega_c(n+\frac{1}{2})$ \cite{Landau1930}, with $n\in\mathbb{N}_0$ the Landau level number and $\hbar \omega_c$ the electron cyclotron energy as before. Thus, state $\ket{d}$, the three electron state with two electrons in the s-shell and one electron in a Landau level (see Fig.\ 3(e) of the main text), has a magnetic field dispersion given by:
\begin{align}
\label{Eq_E_Landau}
E_d=E_d^0+\hbar\omega_c\left(n+\frac{1}{2}\right).
\end{align}
The term $E_d^0$ includes both single particle energies and Coulomb interaction energies in state $\ket{d}$. The dispersion of $E_d$ is dominated by the Landau level dispersion and $E_d^0$ is assumed to be constant. Basis states $\ket{a}$ and $\ket{d}$ are coupled by an Auger-like process: an anti-crossing with a particular Landau level appears at the magnetic field at which the energy difference between Landau level and QD-p$_-$-shell equals that between p$_-$- and s-shell of the QD. (For a harmonic confinement potential in a single particle description, this corresponds to the crossing point of the Landau level and the $d_-$-shell of the QD.) At this point the corresponding Auger-like process is energy conserving and state $\ket{a}$ couples to the corresponding Landau level. The matrix element for the coupling to the 0th Landau level has been calculated in Ref.\ \onlinecite{Karrai2003}. Higher Landau levels also have a d-like component and thus couple to state $\ket{a}$ as well \cite{Karrai2003}. This gives rise to a series of anti-crossings in the emission spectrum of the X$^{3-}$ singlet state \cite{Karrai2003,VanHattem2013}. Since the magnetic field regime in which state $\ket{a}$ couples to Landau levels is relatively narrow, we take here just the coupling to the $n=0$ Landau level, and we assume that the $\ket{a} \leftrightarrow \ket{d }$ coupling, parameter $A_L$, to be constant with magnetic field \cite{Karrai2003}. Finally, we introduce a tunnel coupling $t_L$ between the $d_-$-shell and the 0th Landau level of the WL \cite{Govorov2003}. This couples basis states $\ket{b}$ and $\ket{d}$ with matrix element $t_d$ (see Fig.\ \ref{fig:suppl_intro}(a)). Including $t_d$ turned out to be necessary to obtain a good description of the measurement, in particular the X$^{3-}$ singlet emission at $\sim4$ T.

In this model, the Auger-like coupling between $\ket{a}$ and $\ket{b}$ is calculated based on the assumption of a perfect harmonic confinement potential. However, for carriers in higher shells of the QD this is not necessarily a good assumption. Especially in the case of a standard InGaAs QD with WL-states, the potential ``softens" (becomes sub-harmonic) such that the energies of the higher shells are reduced with respect to the harmonic oscillator (see Fig.\ \ref{fig:suppl_intro}(b) for an illustration). To compensate for such an effect, we correct both energy and Auger-like coupling of state $\ket{b}$, the only basis state with a QD-d-shell component. In particular, we add a constant energy term to the energy of state $\ket{b}$ ($E_{b}^{0} + E_{b}^{C} \rightarrow E_{b}^{0} + E_{b}^{C}+ \delta E_{b}$) and scale the coupling between states $\ket{a}$ and $\ket{b}$ with a constant pre-factor ($A_{ab} \rightarrow s_{ab}\cdot A_{ab}$).

The full Hamiltonian describing the final states in the $\ket{a}$, $\ket{b}$, $\ket{c}$, $\ket{d}$ basis becomes:
\begin{align}
\label{Eq_H_final}
\hat{H_{f}}=
\begin{bmatrix}
E_{a}^{0}+E_{a}^{C} & s_{ab}\cdot A_{ab} & 0 & -A_L \\
s_{ab}\cdot A_{ab} & E_{b}^{0}+E_{b}^{C} + \delta E_{b} & \Delta_{bc} & t_d \\
0 & \Delta_{bc} & E_{c}^{0}+E_{c}^{C} & 0 \\
-A_L & t_d & 0 & E_d
\end{bmatrix}
\end{align}
To obtain the energies of all final states after optical emission we diagonalise this Hamiltonian numerically for all different magnetic fields. To calculate the actual energies of the optical emission, the eigenenergies are subtracted from the energy of the initial exciton state, i.e.\ the energy of the X$^{3-}$ before radiative recombination. The initial exciton state is depicted in Fig.\ 3(c) of the main text. It couples optically only to the final state $\ket{a}$. The relative brightness of one particular optical emission line is given by the probability of component $\ket{a}$ in the particular eigenstate of $\hat{H_{f}}$.

A numerical simulation taking also the couplings $A_L$ and $t_L$ to the 0th WL-Landau-level into account is shown in Fig.\ \ref{fig:simulation_explain}(c). We achieve an excellent agreement with the experiment, Fig.\ \ref{fig:simulation_explain}(d). The model accounts for all the main features in the experiment, both the energies of the multiple emission lines and their relative intensities. All the major experimental features are reproduced. This gives us confidence that the model accounts for all the significant interactions in the X$^{3-}$ final state in the most complicated case, an InGaAs QD with associated WL. 

\subsubsection{The X$^{3-}$ initial states}
After X$^{3-}$ recombination, there is a vacancy in the s-shell, allowing Auger-like processes to take place. These processes admix the QD-d-shell and, should they exist, the WL-Landau-levels, to the available final states. The X$^{3-}$ final states have therefore a rich structure. As we argue in the main text, they provide an ideal way of exploring the single particle states at energies well above both the s- and p-shells. The X$^{3-}$ initial states have a much simpler structure.

At high magnetic field, the X$^{3-}$ initial state has two electrons in the conduction band s-shell and two electrons in the p$_-$-shell of the QD along with a hole in the valence band s-shell. For the electrons, this represents a spin singlet. Other configurations have considerably larger single particle energies and are therefore ignored. The energy of this exciton is given by an effective band gap of the QD $E_{g}^{*}$ plus the sum of electron and hole single particle energies and Coulomb interaction terms:
\begin{equation}
\label{Eq_E_initial}
E_{i} = E_{g}^{*} + E_{i}^{0} + E_{i}^{ee} + E_{i}^{eh}.
\end{equation}
Here the term $E_{i}^{0}$ denotes the single particle energy of electrons and hole, $E_{i}^{ee}$ is the sum of the Coulomb interactions between the electrons, and $E_{i}^{eh}$ is the Coulomb interaction between electrons and hole. The hole wave function and its single particle energy can be obtained by using Eqs.\ \ref{Eq_FockD} and \ref{Eq_2D_Psi}, replacing the electron effective mass with the hole effective mass $m_{h}^{*}$. For the single particle energy of electrons and hole in the X$^{3-}$ singlet state we obtain:
\begin{align}
\label{Eq_H_initial_sp}
E_{i}^{0} &= 6\hbar\omega_1-\hbar\omega_c+\hbar\omega_{1}^{h}, \, \text{with}\\
\omega_1^{h}&=\sqrt{\left(\omega_{0}^{h}\right)^2+\left(\frac{\omega_{c}^{h}}{2}\right)^2},
\end{align}
where $\hbar \omega_{c}^{h}=\hbar eB/m_{h}^{*}$ is the cyclotron energy of the hole. We make here the assumption that the QD-confinement-potential experienced by the hole equals that experienced by the electrons. This leads to the relation $m_{h}^{*}\left(\omega_0^{h}\right)^2 = m_{e}^{*}\left(\omega_0\right)^2$ and determines implicitly the parameter $\omega_0^{h}$. For the electron-electron and electron-hole Coulomb energies we obtain \cite{Warburton1998}:
\begin{align}
\label{Eq_E_initial_Coulomb}
E_{i}^{ee} &= \frac{67}{16}\frac{e^2}{4\pi\epsilon_0\epsilon_r}\frac{1}{L_e}\sqrt{\frac{\pi}{2}},\\
E_{i}^{eh} &= -2\frac{e^2\sqrt{\pi}}{4\pi\epsilon_0\epsilon_r}\left(\frac{1}{\sqrt{L_e^2+L_h^2}}+\frac{2L_e^2+L_h^2}{2\left(L_e^2+L_h^2\right)^{3/2}}\right).
\end{align}
Note that the magnetic field dependence of the electron and hole effective lengths, $L_{e}=\sqrt{\hbar/(\omega_1m_{e}^{*})}$ and $L_{h}=\sqrt{\hbar/(\omega_{1}^{h}m_{h}^{*})}$, causes a magnetic field dependence of the electron-electron and electron-hole Coulomb matrix elements, leading in turn to a magnetic field dependence of the emission energies beyond the diamagnetic shift in a single particle picture. In our simulations this effect is taken into account.

There are two final points. First, there is a clear Zeeman effect in the experimental data (Fig.\ \ref{fig:simulation_explain}(d)). The Zeeman energy $E_Z=g_X\mu_BB$, where $g_X=g_e+g_h$ is the exciton g-factor, is included after computing the eigenstates of Eq.\ \ref{Eq_H_final}. The Zeeman effect splits every emission line into two; the energy separation of the lines is the Zeeman energy. This holds for negligible spin-orbit interaction such that spin and orbital degrees of freedom can be considered separately. Secondly, at low magnetic fields, the X$^{3-}$ initial state is an electronic triplet state: the s-shell is doubly occupied, the p$_+$-shell is singly occupied, and the p$_-$-shell is singly occupied. This triplet initial state results in two PL-lines \cite{Warburton2000,Karrai2003}. The triplet state is a less sensitive probe to higher lying single particle states: it does not show a hybridization with the d-shell or with WL-Landau-levels \cite{Karrai2003}. However, at small magnetic fields, around $\sim 1.3$ T, the singlet initial state becomes the ground state, not the triplet state. As such, we have focussed the entire calculation on the X$^{3-}$ singlet initial state. To include the X$^{3-}$ triplet initial state in the simulations, we include it phenomenologically with a parabolic dispersion. 

\subsubsection{Parameters}
The  parameters for the simulations shown in Fig.\ \ref{fig:simulation_explain} and Fig.\ 3(a),(b) of the main text are given in Table \ref{table_sim_X3m}. For both the standard InGaAs QD, and the AlAs-capped QD, the model parameters were tuned to give a quantitative description of the experimental results.

\begin{table}[tb]
\begin{ruledtabular}
\begin{tabular}{lcc} 
& standard InGaAs QD & AlAs-capped QD \\\cline{2-3} 
$\hbar\omega_0\ (\text{meV})$ & 24.163 & 27.524\\
$\hbar\omega_0^h\ (\text{meV})$ & 9.125 & 10.653\\ 
$m_{e}^{*}/m_o$ & 0.0727 & 0.0750\\
$m_{h}^{*}/m_o$ & 1.0 & 0.501\\
$s_{ab}$ & 0.443 & 0.921\\
$s_{ab}\cdot A_{ab}\ (\text{meV})$ & -2.038 & -4.677\\
$E_g^* (\text{eV})$ & 1.3102 & 1.2993\\
$\Delta_{bc}\ (\text{meV})$ & 0.774 & 1.033\\
$t_d\ (\text{meV})$ & 1.549 & --\\
$A_L\ (\text{meV})$ & 0.383 & --\\
$E_d^0\ (\text{meV})$  & 154.2 & --\\  
$\delta E_{b}\ (\text{meV})$ & -4.639 & 10.372\\
$\epsilon_r$ & 13.16 & 12.93\\
$g_X$ & 1.52 & 1.60\\
\end{tabular}
\caption{\label{table_sim_X3m}Parameters for the simulation of the X$^{3-}$ singlet emission lines. Simulation and data for a standard InGaAs and an AlAs-capped QD are shown in Fig.\ \ref{fig:simulation_explain} and Fig.\ 3(a),(b) of the main text. Parameter values are stated with high precision to facilitate reproducing the simulations.}
\end{ruledtabular}
\end{table}

For the AlAs-capped QDs, there is no evidence whatsoever in the PL-spectra for the process related to the hybridization with the WL. This is evidence that the WL for electrons no longer exists. The terms $A_L$ and $t_d$ are therefore set to zero in the simulation. 

For the standard InGaAs QD, the intra-dot Auger-like process (i.e.\  the $\ket{a} \leftrightarrow \ket{b}$ coupling) results in two pairs of emission lines. For the AlAs-capped QD, only a single emission pair is observed. This is evidence that the d-shell of the QD is increased in energy with respect to harmonic confinement. For the simulation, we add a positive correction to the d-shell energy: $\delta E_{b} \sim + 10\ \text{meV}$. Neglecting Coulomb interactions, this leads to a splitting of $\delta E_{b}$ between states $\ket{a}$ and $\ket{b}$. In the X$^{3-}$ final states, the ``second" energy pair of emission lines are red-shifted by $\sim \delta E_{b}+2\cdot s_{ab}\cdot\abs{A_{ab}}\sim20\ \text{meV}$. Furthermore, since $\delta E_{b}>s_{ab}\cdot\abs{A_{ab}}$, the admixture of basis state $\ket{a}$ in the final state corresponding to the ``second" emission pair is strongly reduced: the intensity of the second emission lines becomes very weak.

Additional evidence that the QD-potential ``hardens" (becomes super-harmonic) on capping with AlAs comes from ensemble PL-measurements which show a larger p--d splitting compared to the standard InGaAs QDs (Fig.\ 1(i) of the main text). Microscopically, this behaviour must result from the AlAs which surrounds the QD laterally. 

For InGaAs QDs with a WL, $\delta E_b$ is negative. This is consistent with the concept that the presence of a WL ``softens" the confinement potential at higher energies. However, in comparison to the confinement energy $\hbar\omega_0$ the term $\delta E_b$ is small. In other words, for these standard QDs, the approximation of harmonic confinement is still a reasonable one.

The parameters for the electron and hole confinement energies ($\hbar\omega_0$ and $\hbar\omega_{0}^{h}$) are comparable to literature values \cite{Babinski2006c,Babinski2013}. The term $\hbar\omega_0$ is larger for the AlAs-capped QD. This is consistent with the understanding that the AlAs-capping increases the lateral carrier-confinement of the QD.

The values for the electron effective mass are similar to the bulk effective mass of GaAs \cite{Nakwaski1995} in both cases. In contrast, the hole effective masses are larger than the bulk value. For the calculation shown here, the in-plane effective mass has to be considered and the effective mass is influenced by the strong confinement of the QD \cite{Zhou2009}. Nevertheless, the hole mass is rather large. This may reflect the fact that the relation $m_{h}^{*}\left(\omega_0^{h}\right)^2 = m_{e}^{*}\left(\omega_0\right)^2$ is inaccurate. There is insufficient information in the spectra to determine $m_{h}^{*}$ and $\omega_0^{h}$ independently.

Values obtained for the dielectric constant $\epsilon_r$ are in both cases between the bulk value of GaAs and InAs ($\epsilon_\text{GaAs}=12.5$ \cite{Strzalkowski1976,Grundmann1995}, $\epsilon_\text{InAs}=15.2$ \cite{Grundmann1995}).

\subsubsection{Coulomb Matrix Elements for 2D Harmonic Oscillator Wavefunctions}
\label{sec:CoulomnMatrix}
In the previous section, the direct Coulomb interaction and the exchange interaction between particles is calculated. In this section we give a brief description how these energies can be calculated analytically.

Matrix elements are calculated by inserting the single particle wave functions (see Eq.\ \ref{Eq_2D_Psi}) into Eqs.\ \ref{Eq_direct_Coulomb} and \ref{Eq_exchange_Coulomb}. One obtains a sum of integrals wich have the form:
\begin{gather}
\label{Eq_first_integral}
\iiiint_{-\infty}^{\infty}\frac{e^{-\frac{\alpha}{2}\left(x_1^2+y_1^2+x_2^2+y_2^2\right)}}{\sqrt{(x_1-x_2)^2+(y_1-y_2)^2}} x_1^{n_1}y_1^{n_2}x_2^{n_3}y_2^{n_4}\ dx_1\ dy_1\ dx_1\ dy_2
\end{gather}
After a coordinate transformation into centre of mass coordinates,
\begin{gather}
\label{Eq_center_mass}
X = \frac{1}{2}\left(x_1+x_2\right), Y = \frac{1}{2}\left(y_1+y_2\right), x = \frac{1}{2}\left(x_1-x_2\right), y = \frac{1}{2}\left(y_1-y_2\right),
\end{gather}
an analytical solution for the matrix elements can be obtained by using the integral relation (for even $n,m,N,M$):
\begin{align}
\label{Eq_nmNM}
&\iiiint_{-\infty}^{\infty}\frac{e^{-\alpha\left(x^2+X^2+y^2+Y^2\right)}}{\sqrt{x^2+y^2}} x^ny^mX^NY^M\ dx\ dy\ dX\ dY\nonumber\\
=&\left(\frac{1}{\sqrt{\alpha}}\right)^{N+M+2}\Gamma\left(\frac{N+1}{2}\right)\Gamma\left(\frac{M+1}{2}\right)\int_{0}^{2\pi}\int_{0}^{\infty}e^{-\alpha r^2}r^{n+m}\sin^n\left(\phi\right)\cos^m\left(\phi\right)\ dr\ d\phi\nonumber\\
=&\frac{1}{2}\left(\frac{1}{\sqrt{\alpha}}\right)^{N+M+n+m+3}\Gamma\left(\frac{N+1}{2}\right)\Gamma\left(\frac{M+1}{2}\right)\Gamma\left(\frac{n+m+1}{2}\right)\int_{0}^{2\pi}\sin^n\left(\phi\right)\cos^m\left(\phi\right)\ d\phi.
\end{align}
Here, we used a transformation into polar coordinates and the following relation:
\begin{equation}
\label{Eq_GammaInt_relation}
\int_{0}^{\infty} e^{-\alpha x^2}x^n dx = \frac{1}{2}\left(\frac{1}{\sqrt{\alpha}}\right)^{n+1}\cdot \Gamma\left(\frac{n+1}{2}\right).
\end{equation}
For arbitrary wave functions of a two-dimensional harmonic oscillator, a completely general analytical solution for the Coulomb matrix elements is given in Refs.\ \onlinecite{Wojs1995,Jacak1998}, in agreement with values for specific matrix elements obtained by calculating the integrals one-by-one \cite{Warburton1998}.

\subsection{Wetting-Layer-PL and Indirect Excitons}
\label{sec:PLcomplete}

\begin{figure}[tb]
\includegraphics[width=0.9\columnwidth]{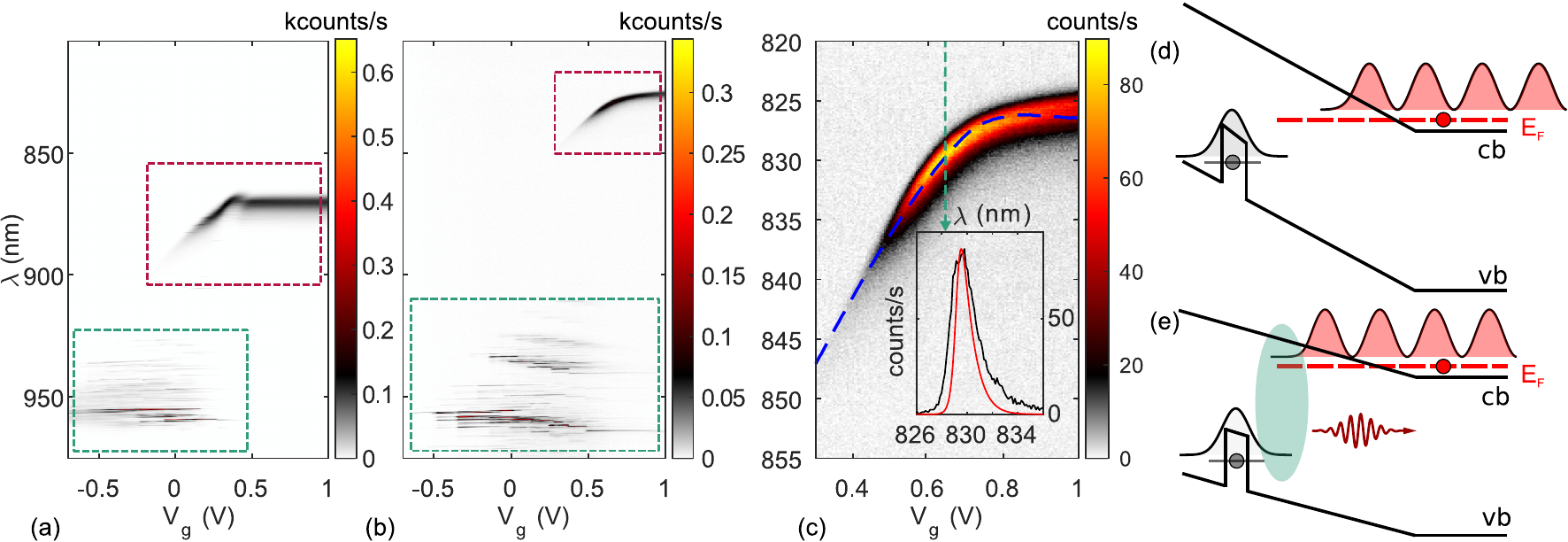}
\caption{(a) PL over a large bandwidth from the reference sample with standard InGaAs QDs. QD-emission appears at high wavelength (green frame). PL at $\sim 875\ \text{nm}$ is related to the WL (red frame). At low gate voltage, the strong shifts with voltage demonstrate that recombination takes place between holes in the WL and electrons in the back-gate \cite{Kleemans2010}. Above $0.45\ \text{V}$, the strong shift with voltage disappears demonstrating that the WL becomes charged with electrons and PL takes place from electrons and holes both confined to the WL. (b) PL over a large bandwidth from the sample with AlAs-capped QDs. The green frame shows the QD-emission-lines. The PL at $\sim830\ \text{nm}$ arises from a spatially indirect recombination between holes in the remnant WL and electrons in the back gate \cite{Kleemans2010,Rai2016}. (c) A simulation of the indirect emission based on a numerical band structure calculation gives reasonable agreement with the experimental data (dashed blue line). Also the asymmetric lineshape of the indirect transition (insert: black curve) is reproduced by the simulation (insert: red curve). (d) Explanation of the indirect emission process: at low gate voltage the in-built electric field of the n-i-Schottky diode suppresses an overlap between back-gate electrons and holes confined in the remnant WL. (e) At positive gate voltage, the electric field is reduced and the electron wave functions from the back gate expand further into the intrinsic region. This leads to a finite wave function overlap between back-gate electrons and WL-holes. The result is an indirect recombination that strongly shifts with the applied gate voltage.}
\label{fig:PLVg_full}
\end{figure}

The nature of the WL-states can also be probed by measuring the emission not from the QDs but from the WL itself. We glean understanding from the standard sample with WL. We then apply this understanding to the sample with AlAs-capped QDs.

Fig.\ \ref{fig:PLVg_full}(a) shows the PL from a sample with standard InGaAs QDs in a large spectral range. QD-emission has a wavelength of about $960\ \text{nm}$; emission related to the WL appears at $\sim 875\ \text{nm}$. At a gate voltage above $0.45\ \text{V}$, the WL is charged with electrons from the back-gate. Optical recombination takes place between an electron and a hole, both in the WL. The quantum confined Stark shift is small because of the strong confinement in the growth direction. At lower gate voltages, the WL-emission strongly shifts with electric field. This signals a change in the emission process: it can be explained with a spatially indirect recombination between an optically generated hole in the WL and an electron from the back-gate \cite{Kleemans2010,Rai2016}.

Fig.\ \ref{fig:PLVg_full}(b) shows PL from the sample with AlAs-capped QDs. The QD-emission has a wavelength of $960\ \text{nm}$, as for QDs in the standard sample. At slightly lower wavelength, emission between higher lying shells of the QD appears. In comparison to the standard InGaAs QDs, this emission is more pronounced for the AlAs-capped QDs. This agrees well with the picture of a stronger confinement caused by the AlAs-capping. Significantly, the emission from a WL-continuum at $\sim875\ \text{nm}$ is not observed for the AlAs-capped QDs. This is confirmation that the electronic WL-states no longer exist. Instead, at high positive gate voltage there is an emission at $\sim830\ \text{nm}$. The emission arises due to a spatially indirect recombination between holes bound by a shallow valence band confinement in the remnant WL and electrons from the back-gate. This process is illustrated in Fig.\ \ref{fig:PLVg_full}(d, e). At small gate voltages, electron wave functions in the back-gate have a negligible overlap with holes bound to the remnant WL.

However, at high positive gate voltages, the electric field is reduced and there is an overlap between electron wave functions in the back-gate and holes in the remnant WL. This gives rise to the emission at $\sim830\ \text{nm}$. The emission shifts strongly with electric field: this proves its spatially indirect nature.

We simulate the indirect emission line from the AlAs-capped sample in a gate voltage range of $0.3-1.0\ \text{V}$. Thereby, we compute the band structure numerically as a function of gate voltage. For the obtained band structure and a given electron energy in the back-gate, we solve the Schr\"odinger equation numerically assuming an infinitely extended back-gate. As a result we obtain the wavefunction $\Psi_k(E)$ of a back gate electron with a specific energy. The intensity of the indirect transition is then estimated as an overlap integral between $\Psi_k(E)$ and the hole wave function $\Psi_h$ which is approximated as a $\delta$-distribution at the WL-position:
\begin{equation}
\label{Eq_indir_Psi_overlap}
I\left(E,T\right)\propto\abs{\langle\Psi_h\mid\Psi_k(E)\rangle}^2\times\text{DOS}\left(E\right)\times f\left(E,T\right).
\end{equation}
The intensity is weighted with a three-dimensional density of states $\text{DOS}\left(E\right)\propto\sqrt{E-E_{\text{cb}}}$, where $E_{\text{cb}}$ is the energy of the conduction band edge, and the Fermi distribution $f\left(E,T\right)$. The best agreement with the data we find for a Schottky barrier of $0.8\ \text{V}$ and a hole ionization energy of $19\ \text{meV}$ ($T=4.2$ K). A slight reduction of the emission energy due to an image charge effect between hole and back gate has been taken into account. The simulation gives a good description of the line position and the asymmetric line shape of the indirect transition (Fig.\ \ref{fig:PLVg_full}(c)).

\subsection{Temperature dependent measurements}
\label{sec:supp_Tdep}
Temperature dependent linewidth measurements in photoluminescence (PL) and resonance fluorescence are shown in Fig.\ 4(a),(b) of the main text. The linewidths (FWHM) as a function of temperature are fitted to the model:
\begin{equation}
\label{Eq_FWHM_T}
\text{FWHM}(T)=\text{FWHM}(T=0)+a\cdot\left[\coth\left(\frac{\hbar\omega_q}{k_{\text{B}}T}\right)-1\right].
\end{equation}
This is the model from Ref.\ \onlinecite{Rudin2006} plus a constant offset. The temperature dependence describes linewidth broadening of the zero-phonon-line by the interaction with a single acoustic phonon mode. The fit parameters for the PL-measurements are shown in Table \ref{table_FWHM_T}.

\begin{table}[t]
\begin{ruledtabular}
\begin{tabular}{lcccccc} 
& X$^{1-}$ & X$^{1-}$ (RF) & X$^{2-}$ & X$^{3-}$ & X$^{4-}$ & X$^{5-}$ \\\cline{2-7}  
$\text{FWHM}(T=0)\ (\mu\text{eV})$  & \textless10 & 2.46 & 11.7 & 35.1 & 39.3 & 169\\  
$a\ (\mu\text{eV})$ & 268 & 29.6 & 182 & 60.1 & 52.3 & 226\\
$\hbar\omega_q\ (\text{meV})$ & 5.94 & 3.14 & 3.78 & 2.10 & 1.55 & 3.01\\
\end{tabular}
\caption{\label{table_FWHM_T}Parameters obtained by fitting Eq.\ \ref{Eq_FWHM_T} to the temperature dependent photoluminescence/resonance fluorescence. Data and fits are shown in Fig.\ 4(a) of the main text.}
\end{ruledtabular}
\end{table}

\subsection{All charge states of a single quantum dot}
\label{sec:full_charge_plateau}
\begin{figure*}[b]
\includegraphics[width=0.9\columnwidth]{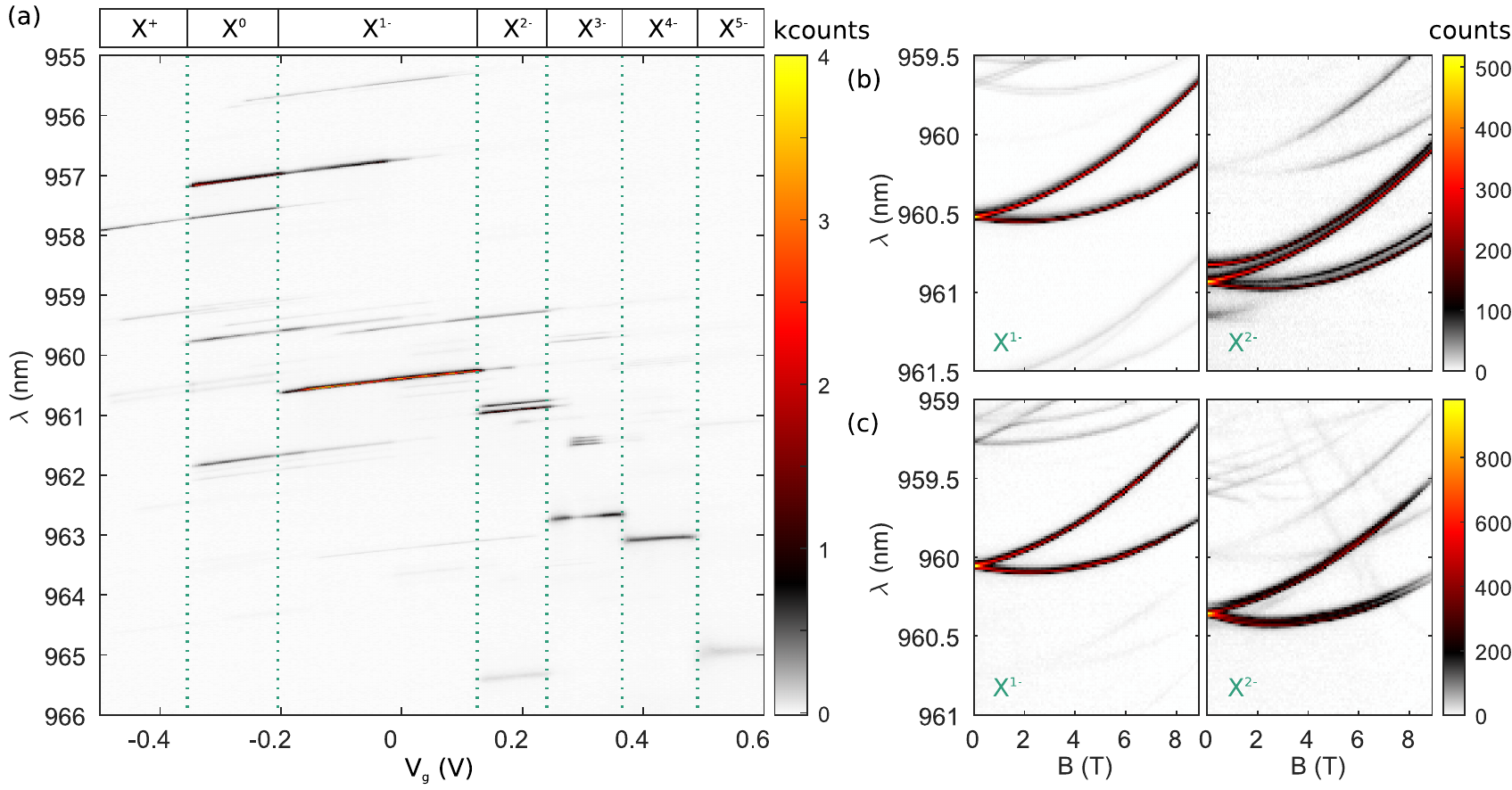}
\caption{(a) Photoluminescence  as function of gate voltage for an AlAs-capped QD without WL-states for electrons. Exciton complexes ranging from X$^+$ to X$^{5-}$ are observed. (b) Emission of an AlAs-capped QD as function of magnetic field (X$^{1-}$ and X$^{2-}$). (c) As for (b) but from a standard InGaAs QD.}
\label{fig:VgFull}
\centering
\end{figure*}
In Fig.\ 2 of the main text, emission of a QD without WL-states for electrons is compared to a standard InGaAs QD with a focus on charged excitons. In Fig.\ \ref{fig:VgFull}(a) we show photoluminescence over a larger gate voltage range (same QD as shown in Fig.\ 2(c)). The measurement shows exciton complexes ranging from a positively charged exciton (X$^+$) to a five-fold negatively charged exciton (X$^{5-}$). As pointed out in the main text, we find optically narrow emission for the excitons X$^{3-}$, X$^{4-}$ and X$^{5-}$ which is in strong contrast to standard InGaAs QDs with WL-states for electrons \cite{Warburton2000}.

Shown in Fig.\ \ref{fig:VgFull}(b) is the photoluminescence of the X$^{1-}$ and the X$^{2-}$ excitons as a function of magnetic field on the same AlAs-capped QD. Both emission lines show a Zeeman splitting and a diamagnetic shift. An equivalent measurement on a standard InGaAs QD is shown in Fig.\ \ref{fig:VgFull}(c) for comparison. The diamagnetic shift and Zeeman splitting are similar for both types of QDs. The X$^{2-}$ emission shows a larger fine structure in the case of the AlAs-capped QD (similar to the X$^{3-}$ triplet emission shown in the main text) implying that the electron-hole exchange in the initial exciton state is larger for the AlAs-capped QDs \cite{Urbaszek2004}.

\end{document}